\documentclass[12pt,preprint]{aastex}

\slugcomment{Not to appear in Nonlearned J., 45.}

\shorttitle{Collapse and Neutrino Emission of Pop III Massive Stars}
\shortauthors{Nakazato, Sumiyoshi, \& Yamada}

\begin{document}


\title{ Gravitational Collapse and Neutrino Emission \\
        of Population III Massive Stars}


\author{Ken'ichiro Nakazato\altaffilmark{1}, Kohsuke
Sumiyoshi\altaffilmark{2,3} and Shoichi Yamada\altaffilmark{1,4}}

\email{nakazato@heap.phys.waseda.ac.jp}

%
%
%
%

\altaffiltext{1}{Department of Physics,
Waseda University, 3-4-1 Okubo, Shinjuku, Tokyo 169-8555, Japan}
\altaffiltext{2}{Numazu Collage of Technology, Ooka 3600, Numazu,
Shizuoka 410-8501, Japan}
\altaffiltext{3}{Division of Theoretical Astronomy, National
Astronomical Observatory of Japan, 2-21-1 Osawa, Mitaka, Tokyo 181-8588,
Japan}
\altaffiltext{4}{Advanced Research Institute for Science \& Engineering, 
Waseda University, 3-4-1 Okubo, Shinjuku, Tokyo 169-8555, Japan}


\begin{abstract}

Population III (Pop III) stars are the first stars in the universe. They
do not contain metals and their formation and evolution may be different
from that of stars of later generations. In fact, according to the
theory of star formation, Pop III stars might have very massive components 
($\sim 100 - 10000M_\odot$). In this paper, we compute the spherically
symmetric gravitational collapse of these Pop III massive stars. We
solve the general relativistic hydrodynamics and neutrino transfer
equations simultaneously, treating neutrino reactions in detail. Unlike
supermassive stars ($\gtrsim 10^5 M_\odot$), the stars of concern in this
paper become opaque to neutrinos.
The collapse is simulated until after an apparent horizon is formed. We
confirm that the neutrino transfer plays a crucial role in the dynamics
of gravitational collapse, and find also that the $\beta$-equilibration
leads to a somewhat unfamiliar evolution of electron fraction. 
Contrary to the naive expectation, the neutrino spectrum does not
become harder for more massive stars. This is mainly because the
neutrino cooling is more efficient and the outer core is more massive as
the stellar mass increases. Here the outer core is the outer part of the
iron core falling supersonically. We also evaluate the flux of relic
neutrino from Pop III massive stars. As expected, the detection of
these neutrinos is difficult for the currently operating
detectors. However, if ever observed, the spectrum will enable us to
obtain the information on the formation history of Pop III stars. We
investigate 18 models covering the mass range of $300 - 10^4 M_\odot$,
making this study the most detailed numerical exploration of spherical
gravitational collapse of Pop III massive stars. This will also serve as
an important foundation for multi-dimensional investigations.   

\end{abstract}





\section{Introduction}

Population III (Pop III) stars are the first stars formed in the
universe. Recently, lots of observational information on them have been
obtained from WMAP (e.g., Spergel et al. 2003),
ground-based and airborne observations of very metal-poor stars 
(e.g., Frebel et al. 2005), and so on. The study of Pop III stars is
indispensable for the understanding of the star formation history of the
universe and the galactic chemical evolution. On the other hand,
theoretical studies of the first star formation suggest that Pop III
stars may have a large population of very massive
objects from hundreds to thousands solar mass (e.g., Nakamura \& Umemura
2001) although their initial mass function (IMF) and star formation rate
are still uncertain.

The evolution and fate of Pop III massive stars have been studied for a
long time (e.g., Bond, Arnett, \& Carr 1984 ; Fryer, Woosley, \& Heger
2001). According to 
these studies, Pop III massive stars lose little of their mass during
the quasi-static evolutions because of zero-metallicity. If an initial
stellar mass is larger than $\sim100M_\odot$, the temperature and
density are such that electron-positron pairs are created copiously at
the center of the star, after the central helium depletion. This process
consumes a part of thermal energy for the 
electron-positron rest mass, making the star unstable against
gravitational collapse. It is called the pair-instability.
The stellar collapse induced by the pair-instability increases the
density and temperature and ignites oxygen. If the stellar mass is less
than $\sim 260M_\odot$, the rapid nuclear burning reverses 
the collapse and disrupts a whole star, giving rise to the so-called
pair-instability supernova. On the other hand, more massive stars
can not halt the collapse by nuclear burnings and emit a large amount of
neutrinos before
forming a black hole. These black holes will eventually have the
mass of their progenitor. It is noted that the mass range is consistent
with that of intermediate mass black holes, whose existence is recently
suggested by observations (e.g., Maillard et al. 2004). Their origin is
unknown at present.

So far, there have been a few papers studying numerically the
gravitational collapse of Pop III massive stars and the emission of
neutrino accompanying it (Fryer, Woosley, \& Heger 2001 ; Iocco et
al. 2005). These studies are imperfect, though, in the treatment of neutrino
transport and the estimate of neutrino spectrum as well as in the
judgment of the black hole formation. On the other hand, there have
been some detailed study of supermassive star collapse, in which
neutrino emissions are taken into account together with general
relativity (Linke et al. 2001). The progenitors of their concern
are probably too massive (over $10^5 M_\odot$) for Pop III stars. It
should be emphasized that the stars of our interest become opaque to
neutrinos before black hole formation unlike the supermassive stars and
that the neutrino transport should be treated appropriately.

In this paper, we calculate 18 models for the spherical gravitational
collapse of Pop III massive stars with different initial masses from
$300M_\odot$ to $13500M_\odot$ and numerically estimate the relic neutrino
flux from them. General relativity and
neutrino physics are treated rigorously. Hence the estimate of neutrino
emission is accurate even when the black hole is formed. We pay
particular attention to the initial mass dependence of the results.

This paper is organized as follows. In
section~\ref{mdlmthds}, we describe the initial models and 
the numerical methods. In section~\ref{collapse}, we present the
numerical results with analysis. We find
some evolutionary features different from what we see in the collapse of
ordinary massive stars producing supernovae.
We also discuss the luminosity and spectrum of emitted
neutrinos and their dependence on the initial stellar mass. In
section~\ref{neutrino}, we evaluate the relic
neutrino flux from Pop III massive stars based on our results and the IMF
recently suggested for Pop III stars. We then consider the insight into
the Pop III star formation history we would obtain from a future 
detection of these neutrino fluxes.

\section{Initial Models and Numerical Methods} \label{mdlmthds}


\subsection{Initial Models}

Pop III massive stars of $\sim100M_\odot$ to $\sim16000M_\odot$ are
supposed to start collapse by the pair-instability. In particular for
stars with the mass $\gtrsim260M_\odot$,
heavy elements are burnt consecutively to form an iron core at the
center. The gravitational collapse is not halted by these energy
generations and a black hole is eventually formed. It is suggested
(Bond, Arnett, \& Carr 1984) that the iron core formed during the
collapse is isentropic and the entropy per baryon is determined by the
oxygen core mass. Although the fraction of helium 
burnt to oxygen in the helium core is uncertain, recent theoretical
calculations suggest that most of helium is consumed
(Umeda, \& Nomoto 2002). The relation between the initial stellar mass
and the helium core mass is also estimated by Bond, Arnett, \& Carr
(1984). Assuming that $90\%$ of helium becomes oxygen, we obtain
the correspondence of the initial mass to the entropy per baryon of iron
core as shown in Table~\ref{m-s_relation}. In addition, the stars with
an oxygen core whose mass is $\gtrsim 8000M_\odot$ become unstable due
to general relativistic effects before they encounter the
pair-instability (Bond, Arnett, \& Carr 1984). In our analysis, the
oxygen core with $8000M_\odot$ corresponds to the initial mass $\sim
16000M_\odot$, the upper limit of the pair-instability given above.
 
In this paper, we start simulations from the iron core formation. We
prepare isentropic iron cores in unstable equilibrium as initial
models. Although the iron core is contracting when formed, we neglect the
motion since the infalling velocity is much smaller than the velocity at
the later phase, which is comparable to the light velocity. We solve the
Oppenheimer-Volkoff equation to obtain the equilibrium configurations.
We set the initial central temperature to be $7.7 \times 10^9$ K, a
typical value for the nuclear statistical equilibrium (NSE). As already
mentioned, we assume that the iron core is isentropic and the entropy
per baryon is given in Table~\ref{m-s_relation} (Bond, Arnett, \& Carr
1984). The subsequent evolution is not so sensitive to this value. We further
assume that the initial electron fraction ($Y_e$) equals to $0.5$ in the
whole core. Hence the entropy per baryon in the iron core is the only
quantity which characterizes 18 models we compare in this paper (see
Table~\ref{m-s_relation}).
The equation of state is adopted from Shen et
al. (1998). The original table of this equation of
state is extended to $T > 100\,\mathrm{MeV}$, since we find this high
temperature in some models. The method of extension is fully consistent
with the original paper (Shen et al. 1998). It is noted that the
equilibrium configurations obtained in this way are unstable to collapse
due to the photodisintegration of iron and as a
result, the core starts to collapse spontaneously without any artificial
reduction of pressure in the initial models.

\subsection{Numerical Methods}

We compute the dynamics of spherical gravitational collapse and the
neutrino transport by the general relativistic implicit Lagrangian
hydrodynamics code, which solves simultaneously the neutrino Boltzmann
equations (Yamada 1997 ; Yamada, Janka, \& Suzuki 1999 ; Sumiyoshi et
al. 2005). This code adopts the metric by Misner \& Sharp (1964),
\begin{equation}
ds^2=e^{2\phi(t,m)}c^2dt^2-e^{2\lambda(t,m)}\displaystyle{\left(\frac{G}{c^2}\right)}^2dm^2-r^2(t,m)\displaystyle{\left(d\theta^2+\sin^2\theta d\phi^2\right)},
\label{metric}
\end{equation}
and solves the evolution of the space-time, where $c$ and $G$ are the
velocity of light and the gravitational constant, respectively, which
are taken to be unity in the following equations, and $t$, $m$ and $r$
are the time coordinate, the baryon mass coordinate and the
circumference radius, respectively. This time-slicing allows us to
follow the dynamics with no difficulty until a black hole
formation. However, the original code is meant to be employed for the 
supernova simulation, in which gravity is not so strong. As the black
hole formation is approached and gravity becomes strong, we find it
better to use the following form for the matter contribution to the
energy equation,  
\begin{equation}
e^{-\phi}\displaystyle\left(\frac{\partial \varepsilon}{\partial t}\right)_\mathrm{m} = -\frac{p}{\Gamma}\frac{\partial}{\partial m}(4\pi r^2U),
\label{yamada8}
\end{equation}
where $\varepsilon$ and $p$ are the specific internal energy and matter
pressure, and $U$ and $\Gamma$ are the radial fluid velocity and
general relativistic Lorentz factor defined as
\begin{equation}
U = e^{-\phi}\frac{\partial r}{\partial t},
\label{yamada11-12}
\end{equation}
\begin{equation}
\Gamma = e^{-\lambda}\frac{\partial r}{\partial m},
\label{yamada15}
\end{equation}
respectively. This form of energy equation is used only when 
$\Gamma < 0.9$.
We use 127 radial mesh points and set
a fixed boundary condition at the outermost grid point. The radius of the
outer boundary is chosen to be large enough not to affect the results.

We compute the neutrino distribution functions for 4 species of neutrino
($\nu_e$, $\bar \nu_e$, $\nu_\mu$, $\bar \nu_\mu$) on a
discretized energy- and angular-grid points by solving the neutrino
Boltzmann equations. In our simulations, the energy space is
discretized to 12 mesh points and the angular space is discretized to 4
mesh points. Using a 1.5 times finer energy mesh, we find a few percent
change in the total neutrino energy etc., 
and the qualitative feature discussed below is unchanged. As already
mentioned, we consider 4 species of neutrino, $\nu_e$, 
$\bar \nu_e$, $\nu_\mu$ and $\bar \nu_\mu$, assuming that  $\nu_\tau$
and $\bar \nu_\tau$ are the same as $\nu_\mu$ and $\bar \nu_\mu$,
respectively. We take into account the following neutrino reactions:\newline
[1] electron-type neutrino absorption on neutrons and its inverse,
\begin{equation}
\nu_e + n  \longleftrightarrow  e + p ,
\label{ecp}
\end{equation}
[2] electron-type anti-neutrino absorption on protons and its inverse,
\begin{equation}
\bar \nu_e + p  \longleftrightarrow  e^+ + n ,
\label{aecp}
\end{equation}
[3] neutrino scattering on nucleons,
\begin{equation}
\nu + N  \longleftrightarrow  \nu + N ,
\label{nsc} 
\end{equation}
[4] neutrino scattering on electrons,
\begin{equation}
\nu + e  \longleftrightarrow  \nu + e ,
\label{esc} 
\end{equation}
[5] electron-type neutrino absorption on nuclei,
\begin{equation}
\nu_e + A  \longleftrightarrow  A + e^- ,
\label{ncp}
\end{equation}
[6] neutrino coherent scattering on nuclei,
\begin{equation}
\nu + A  \longleftrightarrow  \nu + A ,
\label{csc}
\end{equation}
[7] electron-positron pair annihilation and creation,
\begin{equation}
e^- + e^+  \longleftrightarrow  \nu + \bar \nu ,
\label{eppa} 
\end{equation}
[8] plasmon decay and creation,
\begin{equation}
\gamma^\ast  \longleftrightarrow  \nu + \bar \nu ,
\label{plasmon}
\end{equation}
[9] neutrino bremsstrahlung,
\begin{equation}
N + N^\prime \longleftrightarrow N + N^\prime + \nu + \bar \nu .
\label{brems}
\end{equation}
See Sumiyoshi et al. (2005) for the details.




\section{Gravitational Collapse of Pop III Massive Stars} \label{collapse}

\begin{figure}
\plotone{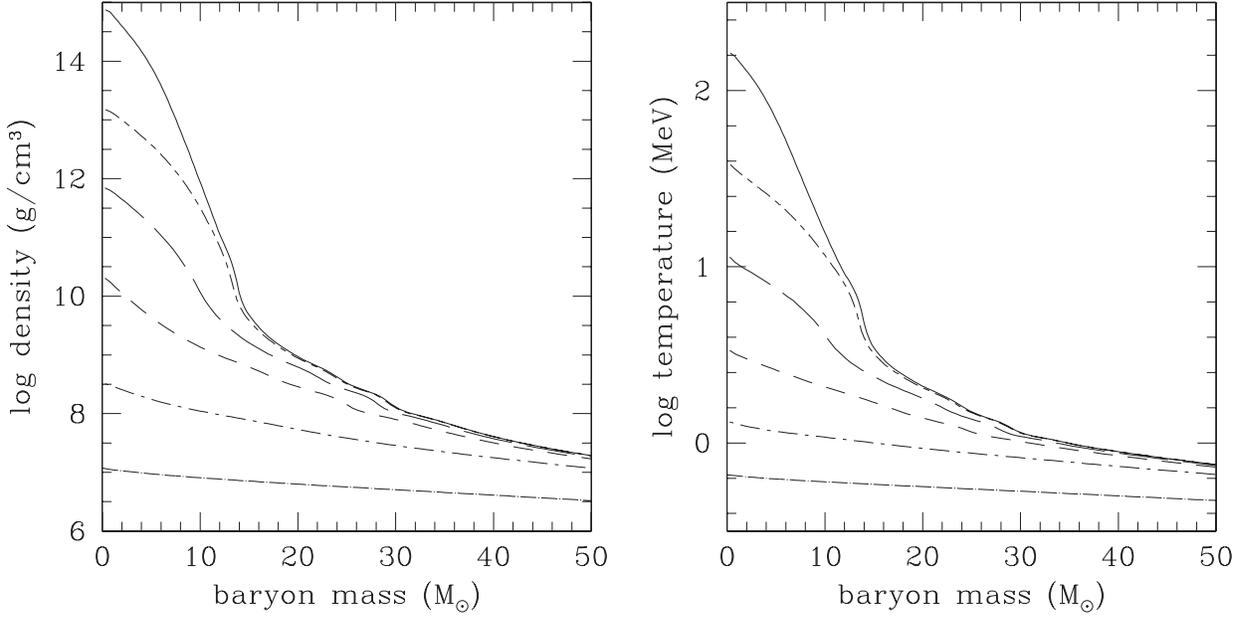}
\caption{Density (left) and temperature (right) profiles for the model with
 $M_i=375M_\odot$. The lines correspond, from bottom to top, to
 $t=-8.88\,\mathrm{s}$, $t=-239\,\mathrm{ms}$,
 $t=-41.6\,\mathrm{ms}$, $t=-12.3\,\mathrm{ms}$,
 $t=-1.52\,\mathrm{ms}$ and $t=0\,\mathrm{ms}$.}
\label{dteg}
\end{figure}

In this section, we summarize the numerical results of the
gravitational collapse. At first we present the result of the reference
model with the initial mass $M_i=375M_{\odot}$, whose density and
temperature profiles are shown in Fig.~\ref{dteg}, and then we make
comparison with different models. In the following, we measure the time
from the point at which the apparent horizon is formed.

\begin{figure}
\plotone{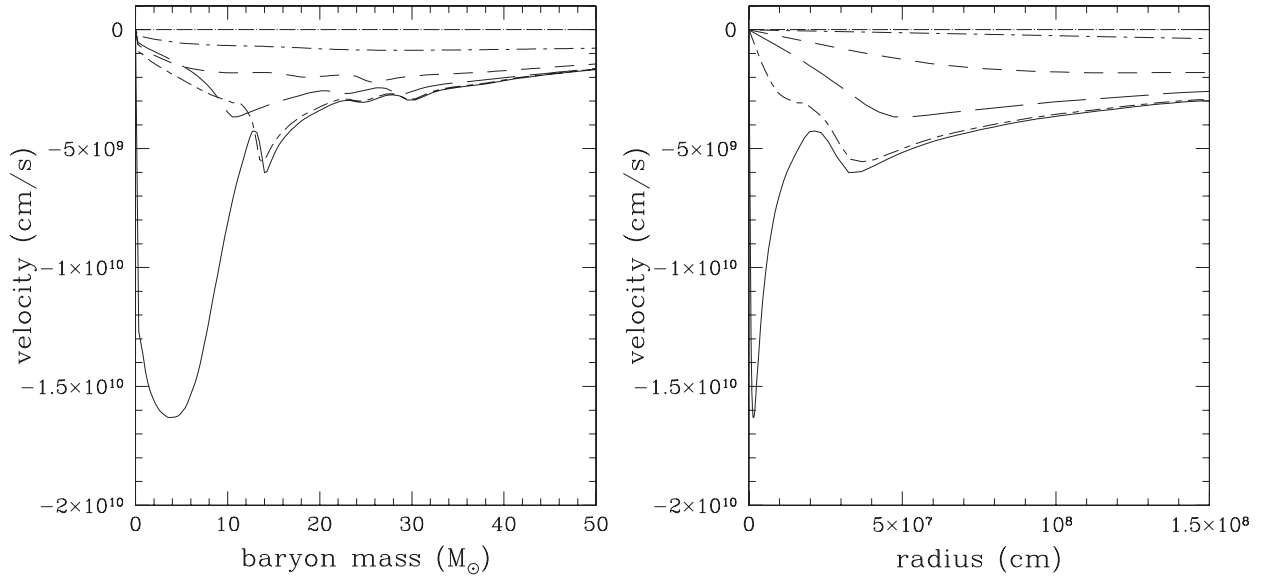}
\caption{Profiles of the radial velocity for the model with
 $M_i=375M_\odot$. The notation of lines is the same as in Fig.~\ref{dteg}.}
\label{velprf}
\end{figure}

\begin{figure}
\plotone{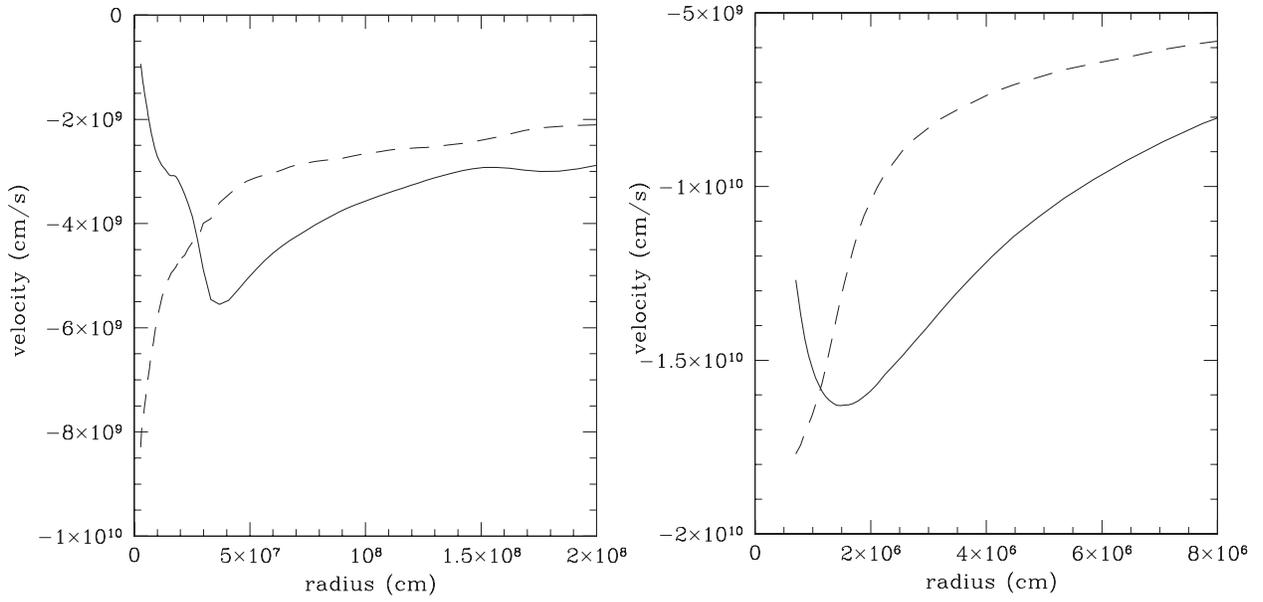}
\caption{Profiles of the radial velocity (solid line) and the sound speed
 (dashed line). The left panel is for $t=-1.52\,\mathrm{ms}$ and
 the right one is for $t=0\,\mathrm{ms}$.}
\label{csvelprf}
\end{figure}

Just like the ordinary collapse-driven supernovae, the collapsing
core is divided into two parts, to which we refer as the inner and
outer cores. The inner core contracts subsonically and homologously 
($U\propto r$) while the outer core infalls supersonically like free-fall
$U\propto r^{-1/2}$, except for the very late phase, in which the black
hole formation is imminent. We can see this structure in
Fig.~\ref{velprf}, in which we plot the radial velocity profiles at
different times for the reference model. In
Fig.~\ref{csvelprf}, we plot also the sound speed profiles at
$t=-1.52\,\mathrm{ms}$ and $t=0\,\mathrm{ms}$ (the time of the apparent
horizon formation) for the same model. Around $t=0\,\mathrm{ms}$, the
homology does not hold any more due to general relativistic strong
gravity and the inner core splits into two parts. It is noted that the
inner core is the main neutrino source especially in the late
phase of collapse. This will be discussed later again.

The collapse described in this paper results in a direct black hole
formation, not experiencing a bounce because of the general relativistic
strong gravity. To judge the black hole formation, we utilize
apparent horizon. The apparent horizon is the outermost trapped
surface. The trapped surface is the surface where both the ingoing and
outgoing null geodesics have a negative expansion. The set of the trapped
surfaces is called a trapped region. In the Misner-Sharp metric, the
condition for the trapped region is expressed as
\begin{equation}
U + \Gamma < 0,
\label{ah}
\end{equation}
which is equivalent to
\begin{equation}
r < r_\mathrm{g} \equiv 2\widetilde m,
\label{ah2}
\end{equation}
where $\widetilde m$ denotes the gravitational mass and it
is related to $U$, $\Gamma$ and $r$ as 
\begin{equation}
\Gamma^2 = 1 + U^2 - \frac{2\widetilde m}{r}.
\label{yamada16}
\end{equation}
Since it is proved that the apparent horizon is always located inside
the event horizon (e.g., Wald 1984), the existence of apparent
horizon is a sufficient condition for the black hole formation.
The apparent horizon forms off center at first, and then it is extended
both inwards and outwards.

\begin{figure}
\plotone{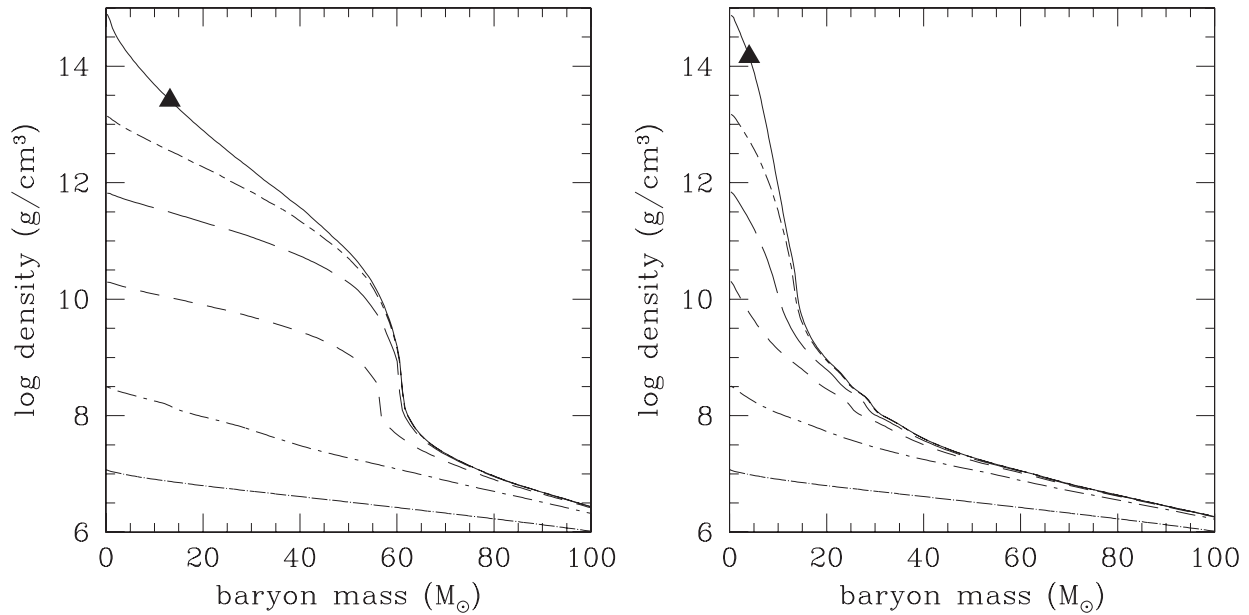}
\caption{Comparison of the results without (left) and with (right) 
 neutrinos for the same model with $M_i=375M_\odot$. In the each 
 panel, the triangles show the locations of the apparent horizon. In
 the left panel, the lines correspond, from bottom to top, to
 $t=-12.0\,\mathrm{s}$, $t=-343\,\mathrm{ms}$,
 $t=-42.3\,\mathrm{ms}$, $t=-4.92\,\mathrm{ms}$,
 $t=-0.699\,\mathrm{ms}$ and $t=0\,\mathrm{ms}$.
 The right panel is the same as the left panel in
 Fig.~\ref{dteg}.}
\label{neueffdns}
\end{figure}

\begin{figure}
\plotone{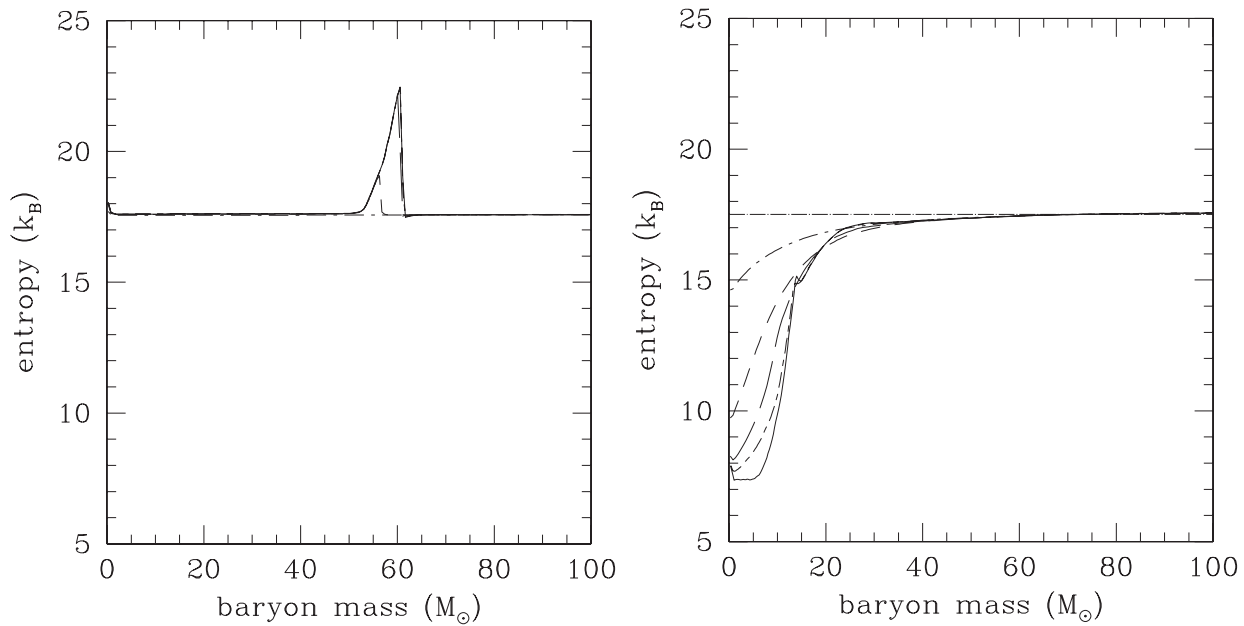}
\caption{The entropy profiles for the model with $M_i=375M_\odot$
 without (left) and with (right) neutrinos. The notation is the same as
 in Fig.~\ref{neueffdns}}
\label{neueffetp}
\end{figure}

In order to emphasize the importance of neutrinos for the dynamics, we
compare the results with and without neutrino transport in
Fig.~\ref{neueffdns}. Since the neutrino cooling is more efficient as
the density increases, the neutrino emission accelerates the collapse in
the central region in the early phase. As a result, the inner core
becomes smaller when the neutrino is taken into account. The formation
of the apparent horizon is also affected. The baryon mass coordinate
where the apparent horizon forms for the first time is $4.08 M_{\odot}$
for the model with neutrino transport, whereas that of the model without
neutrino transport is $13.2 M_{\odot}$. Moreover, the time until the
apparent horizon appears becomes shorter.
In Fig.~\ref{neueffetp}, we show the evolution of the entropy per
baryon. For the model without neutrino transport, the dynamics is
adiabatic and the entropy is constant except at the shock formed around
the core surface ($\sim60M_\odot$), where the entropy is generated. For
the model with neutrino transport, on the other hand, the entropy per
baryon decreases because of the neutrino cooling.

\begin{figure}
\plotone{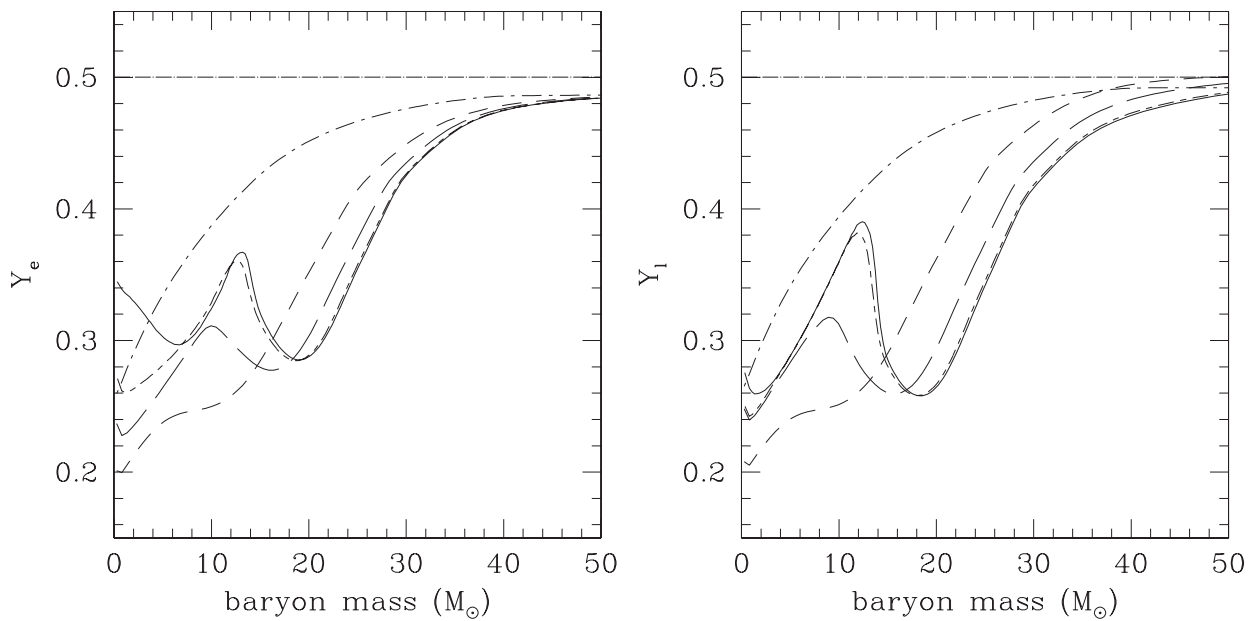}
\caption{Profiles of $Y_e$ (left) and $Y_l$ (right) for the model with
 $M_i=375M_\odot$. The notation of lines is the same as in Fig.~\ref{dteg}.}
\label{yeyl}
\end{figure}

\begin{figure}
\plotone{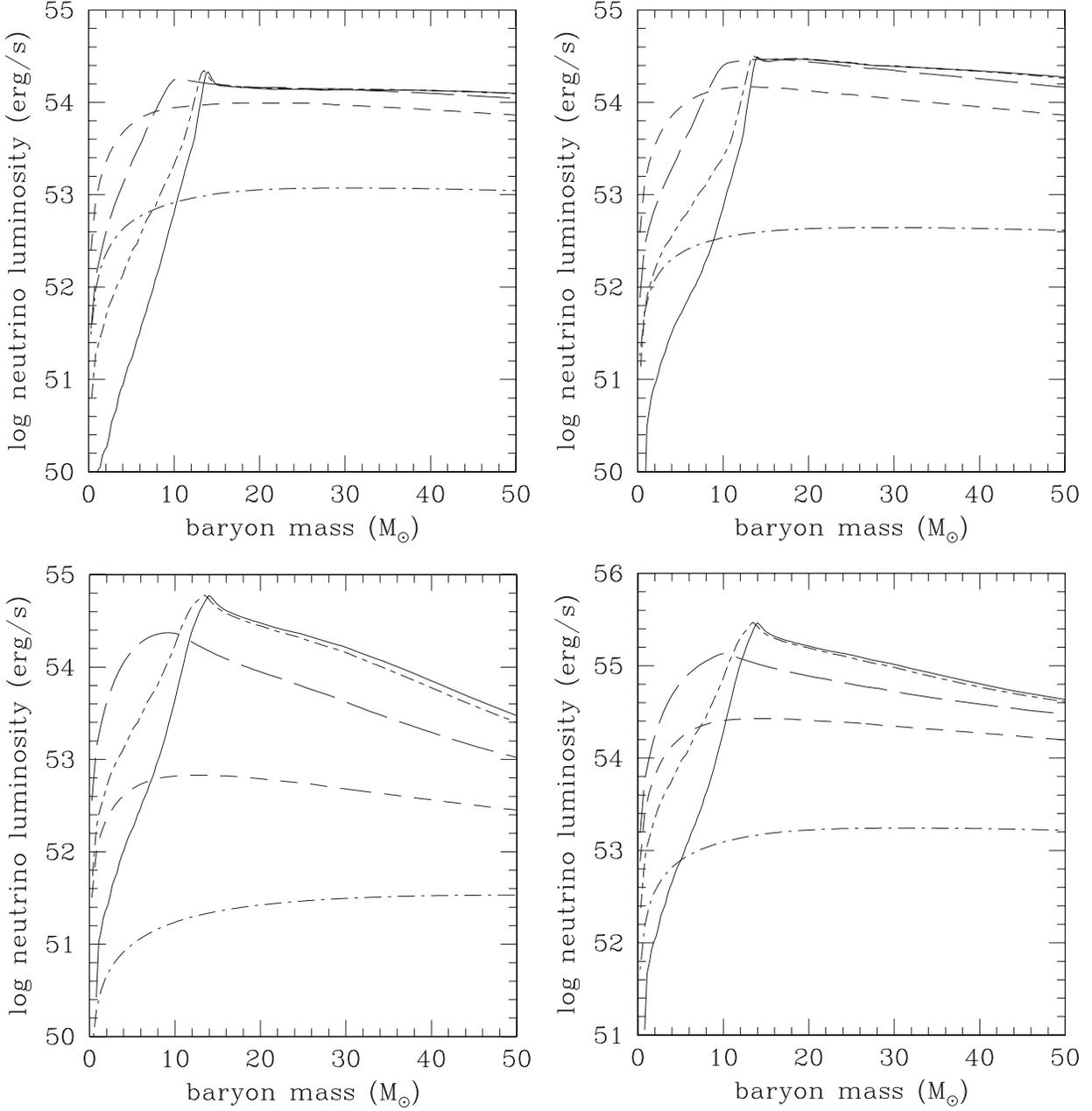}
\caption{Profiles of neutrino luminosity for the model with
 $M_i=375M_\odot$. Upper left, upper right, lower left and lower right
 panels are for $\nu_e$, $\bar \nu_e$, $\nu_x$
 and their sum, respectively, where $\nu_x$ stands for $\mu$- and
 $\tau$-neutrinos 
 and anti-neutrinos. The notation of lines is the same as in
 Fig.~\ref{dteg}.}
\label{neulum}
\end{figure}

We now turn to the evolutions of electron fraction ($Y_e$). As the
electron capture proceeds, $Y_e$ is depleted as shown in
Fig.~\ref{yeyl}. Near the center, however, $Y_e$ then starts to
increase, which is not seen in the ordinary supernova core. The main
reason is that the electrons are not degenerate due to high entropy
in the present model and the reaction rates of electron capture and
positron capture are different (e.g., Bruenn 1985). When the neutrino
energy is lower than or comparable to the mass difference of proton and
neutron, the electron capture dominates over the positron capture
considerably. As the neutrino energy becomes higher than the 
mass difference and electrons are not degenerate, these two reaction
rates become comparable. As a result, the electron
capture is dominant and equilibrated first, and the positron capture
catches up afterward, leading to the rise of $Y_e$ before the complete
$\beta$-equilibrium. In Fig.~\ref{yeyl}, we also show the evolution of
lepton fraction, $Y_l$. Before neutrino trapping, the $Y_l$ profile
evolves in the same way as the $Y_e$ profile. Then, $Y_l$ becomes
unchanged and only $Y_e$ varies after neutrino trapping. We can 
confirm these behaviors in the neutrino luminosities given in
Fig.~\ref{neulum}.

A large amount of neutrinos are emitted by the gravitational collapse of
Pop III massive stars. The luminosity becomes as high as
$\sim10^{54}\,\mathrm{erg/s}$, much greater than the value for the
ordinary supernova $\sim10^{53}\,\mathrm{erg/s}$. However, the total
energy is $\sim10^{53}\,\mathrm{erg}$, which is comparable to the
ordinary supernova. This is because the apparent horizon is formed
during the collapsing phase and, as a result, the neutrino emission
lasts only for $\sim100\,\mathrm{ms}$.

For the stars of current interest, the neutrino sphere is formed in
general before the black hole formation.
It is also noted that the
apparent horizon is formed inside the neutrino sphere. The neutrino
optical depth $d(r)$ at the radius $r$ is defined as
\begin{equation}
d(r) = \int^{R_s}_{r} \frac{dr^{\prime}}{l_\mathrm{mfp}(r^{\prime})},
\label{depth}
\end{equation}
where $R_s$ is the stellar radius and $l_\mathrm{mfp}$ is the mean
free path of neutrino. The radius of neutrino sphere $R_\nu$ is defined
as 
\begin{equation}
d(R_\nu) = \frac{2}{3}.
\label{sphere}
\end{equation}
Because the mean free path depends on the energy and species of
neutrino, so does the neutrino sphere.

\begin{figure}
\plotone{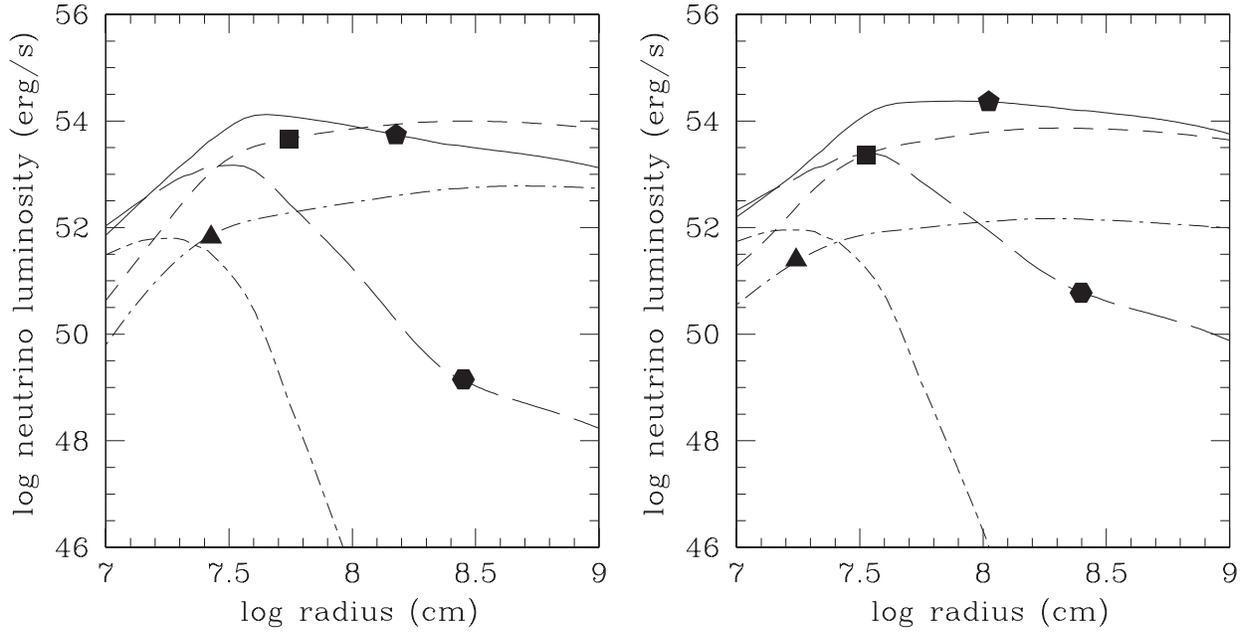}
\caption{Profiles of the neutrino luminosity with different energies for the
 model with $M_i=375M_\odot$ at $t=-12.3\,\mathrm{ms}$. Left and right
 panels are for $\nu_e$ and $\bar \nu_e$, respectively. In each panel,
 the dot-dashed line, short dashed line, solid line, long dashed
 line and short-long dashed line correspond to the neutrino luminosity
 with $2.51\,\mathrm{MeV}$, $6.31\,\mathrm{MeV}$, $15.8\,\mathrm{MeV}$,
 $35.4\,\mathrm{MeV}$ and $70.7\,\mathrm{MeV}$. The points on 
 each line mark the locations of neutrino sphere for each energy.}
\label{dle}
\end{figure}

In Fig.~\ref{dle}, we show the neutrino luminosity profiles and the
location of the neutrino sphere for $\nu_e$ and $\bar \nu_e$ with
several energies at $t=-12.3\,\mathrm{ms}$. At this moment, the inner
core surface
is located at $r=5\times10^7\,\mathrm{cm}$ and we can recognize that the
neutrino luminosity decreases towards the center inside the inner core.
This is because the interactions of neutrinos with matter are frequent
enough to make the neutrino angular distribution isotropic. In the outer
core, neutrinos created inside the inner core are flowing out and the
neutrino angular distribution is not isotropic. In our model, the
outer core is as thick as $\sim10^{9}\,\mathrm{cm}$ and, as a result,
the neutrino sphere is far away from the inner core surface for high
energy neutrinos. For instance, the optical depth of $\nu_e$ with
$15.8\,\mathrm{MeV}$ on the inner core surface is about $4$. These
neutrinos are mainly emitted from the inner core and are absorbed or
scattered by nucleons and electrons. As a result, the luminosity
decreases mildly as neutrinos propagate from the inner core surface to
the neutrino sphere. For the neutrinos with much higher energy, on the
other hand, the reduction of the luminosity
is steeper because the neutrino optical depth is larger, whereas the
neutrino sphere for neutrinos with lower energy is close to or inside
the inner core surface.

The luminosities of low energy neutrinos increase outside the neutrino
spheres. Although these low energy neutrinos have ceased to react with
matter, but the neutrinos with higher energy are still scattered down to
lower energy, raising the luminosities of low energy neutrinos. This is
also the reason why the mean energy of neutrinos is reduced
substantially from the value at the inner core surface
(main neutrino source) as they propagate to the surface. It is even
lower than the mean energy for the ordinary
supernova. Incidentally, in the outermost region (
$\gtrsim3\times10^8\,\mathrm{cm}$, in Fig.~\ref{dle}), the neutrino
luminosities decrease slightly. This reflects the difference of the
emission time, that is, as the collapse goes on, the matter become
denser and hotter and the neutrino emission occurs more efficiently.

\begin{figure}
\plotone{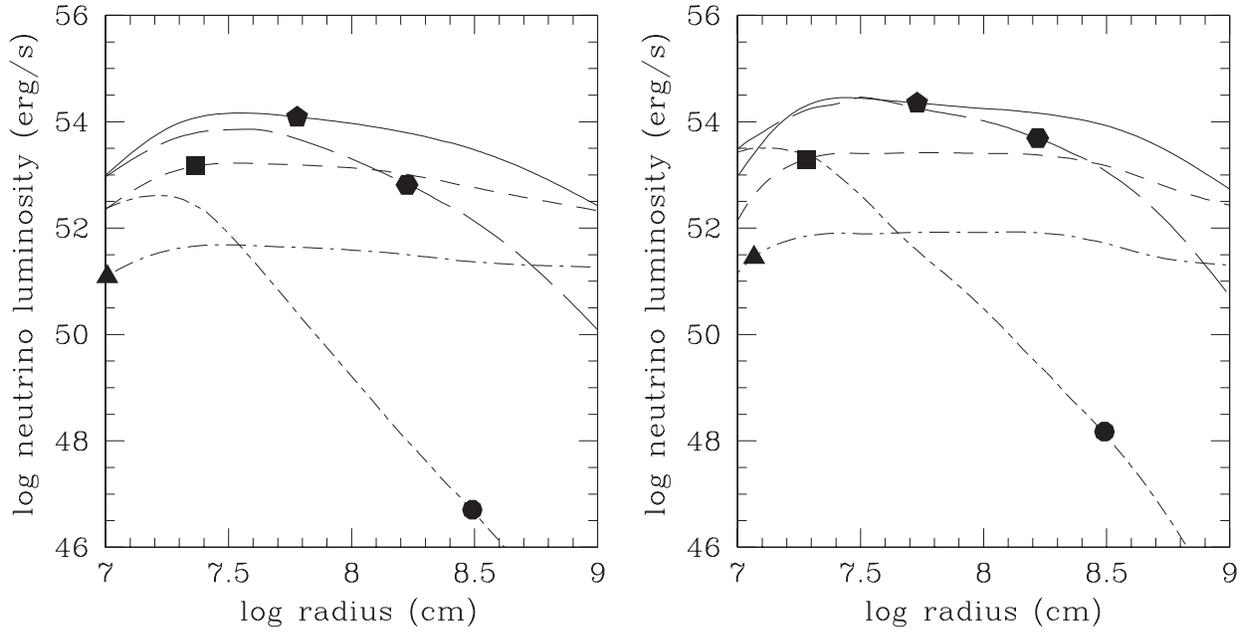}
\caption{Profiles of the $\nu_x$ ($=$ $\nu_\mu$, $\nu_\tau$,
$\bar \nu_\mu$, $\bar \nu_\tau$) luminosity with different energies for
 the model with $M_i=375M_\odot$ at $t=-12.3\,\mathrm{ms}$ (left) and
 $t=-1.52\,\mathrm{ms}$ (right). The notation of lines and points is
 the same as in Fig.~\ref{dle}.}
\label{dlx}
\end{figure}

We also investigate the other neutrino species, namely $\mu$- and 
$\tau$-neutrinos, and anti-neutrinos. In Fig.~\ref{dlx}, we show the
luminosity profiles and the locations of neutrino sphere for $\nu_\mu$
at $t=-12.3\,\mathrm{ms}$ and $t=-1.52\,\mathrm{ms}$. It is noted that
we assume that  $\nu_\tau$ ($\bar \nu_\tau$) is the same as $\nu_\mu$
($\bar \nu_\mu$). The luminosities of $\nu_\mu$ and $\bar \nu_\mu$ are
almost identical because they have the same reactions, the difference of
coupling constants is minor and $Y_e$ is not so small.
In the following, ignoring this tiny difference, we denote these 4
species as $\nu_x$ collectively. For $\nu_x$, the radii of the
neutrino spheres are smaller than those of $\nu_e$ and $\bar \nu_e$,
since $\nu_x$ do not react via charged current processes. 
Outside the neutrino sphere, the negative gradient of the luminosity
is steeper for $\nu_x$ than for $\nu_e$ and $\bar \nu_e$, especially for
high energy neutrinos.
The absence of the charged current reactions makes the core more
optically thin for $\nu_x$, and the rise of the temperature of the inner
core is reflected more immediately by the neutrino luminosity.

\begin{figure}
\plotone{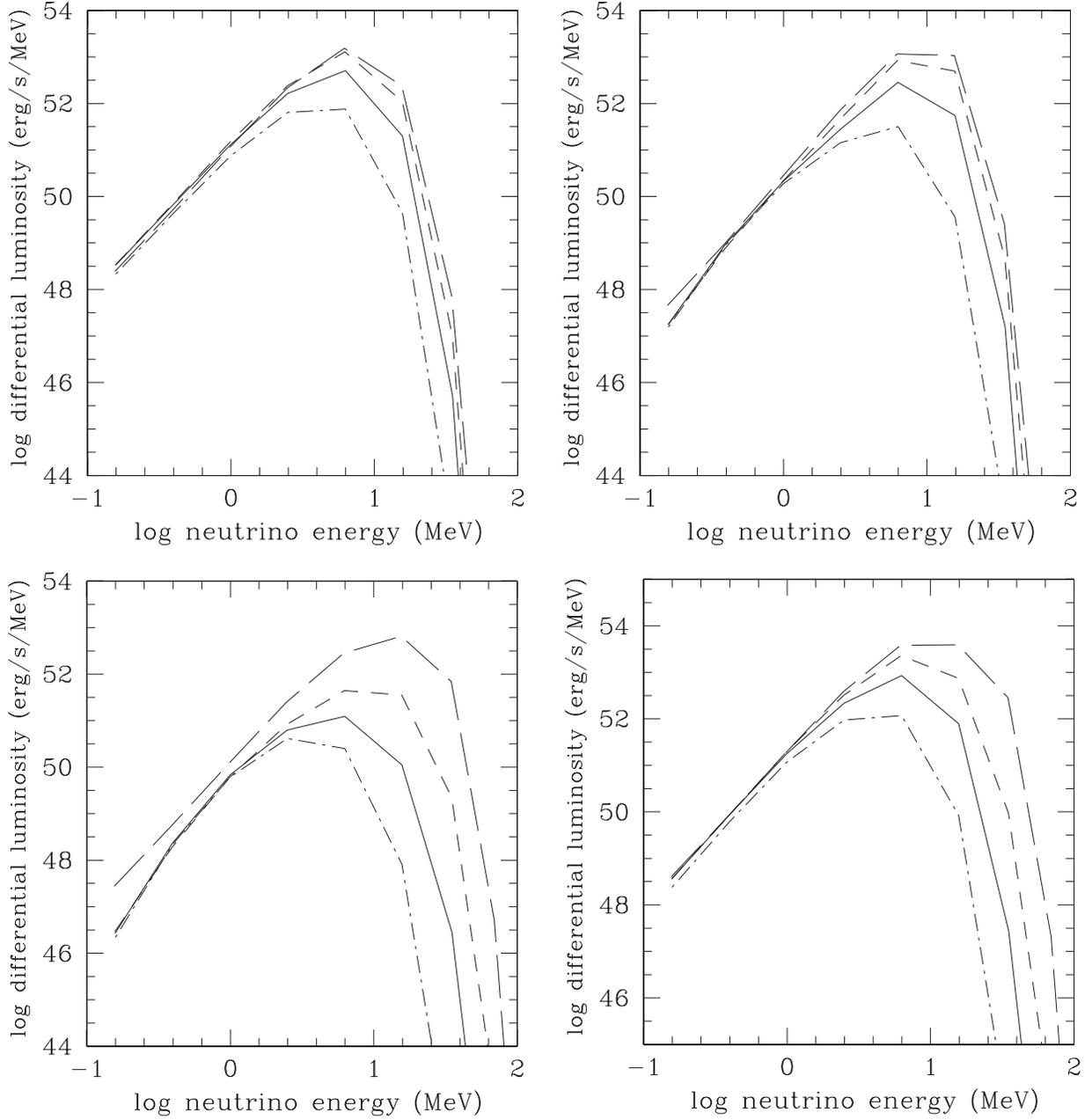}
\caption{Spectra of emitted neutrinos. Upper left, upper right, lower
 left and lower right panels are for $\nu_e$, $\bar \nu_e$, $\nu_x$ ($=$
 $\nu_\mu$, $\nu_\tau$, $\bar \nu_\mu$, $\bar \nu_\tau$)
 and their sum, respectively. The dot-dashed line, solid
 line, short dashed line and long dashed line represent, respectively,
 the spectra at  $t=-239\,\mathrm{ms}$, $t=0\,\mathrm{ms}$,
 $t=60.3\,\mathrm{ms}$ and $t=85.0\,\mathrm{ms}$.}
\label{speev}
\end{figure}

It is already noted that the average
energies of emitted neutrinos are rather lower than those for the
ordinary supernova. The time evolutions of neutrino spectra on the
surface of our computation region for the reference model
are shown in Fig.~\ref{speev}. The spectra after $t=0\,\mathrm{ms}$
are evaluated under the assumption that the neutrinos outside the
neutrino sphere flow freely. The last time in Fig.~\ref{speev},
$t=85.0\,\mathrm{ms}$, is the light crossing time from the neutrino
sphere for high energy neutrinos with appreciable population (roughly
the location of the filled circle in the right panel of Fig.~\ref{dlx})
to the surface. We can see that the neutrino spectra get harder as the
time passes. This is because the temperature of the inner core namely
becomes hotter as the collapse goes on.
 
We compare our results with those of Fryer, Woosley, \& Heger (2001).
Their non-rotation model with an initial mass of $300M_{\odot}$ gives a
black hole at the end. The oxygen core mass is $\sim180M_{\odot}$ (see
Fig.~2 in their paper). We choose the reference model for comparison,
since it has the same oxygen core mass $M_\mathrm{O}\sim180M_{\odot}$,
though the total mass is a little bit larger, $M_i=375M_{\odot}$.

The qualitative features is similar. For example, the collapsing iron
core is divided into two parts, the inner and outer cores, the
electron fraction ($Y_e$) starts to increase near the center after it
decreases at first
and the black hole is formed without the bounce. However, the
quantitative differences can be recognized. The most outstanding one is
the size of the inner core. In our model, it is $\sim14M_{\odot}$ while
it is $\sim40M_{\odot}$ (see Fig.~3 in their paper) in their
model. Although the difference is partially ascribed to the difference
of the initial models, the main reason, we think, is the 
treatment of the neutrino transport. In their model, the neutrinos are
treated in the so-called grey approximation, where only the
energy-averaged distribution is employed. As demonstrated earlier, the
location of neutrino sphere is highly energy-dependent. With only an
average energy taken into account, the neutrino trapping will be
overestimated. On the other hand, we treat multi-energy groups in our
simulations, and the inner core can be cooled by low energy
neutrinos. Due to this enhanced neutrino cooling, the inner core becomes
smaller in our model.

The total emitted energy also differs. Our model estimates it to be
$\sim4\times10^{53}\,\mathrm{erg}$, which is an order of magnitude
smaller than $\sim3\times10^{54}\,\mathrm{erg}$ in their model (see
Fig.~4 in their paper). The discrepancy seems to come from the employed
criterion of black hole formation. They assumed that the black hole is
formed at the time when a certain fraction of the star falls within the
last stable orbit, $r=3r_\mathrm{g}$ (e.g., Shapiro \& Teukolsky 1983),
while we utilized the apparent horizon formation, which is a rigorous
sufficient condition, as mentioned already. This difference leads to the
substantial difference of the duration of neutrino emissions. In fact,
neutrino emissions last for $\sim100\,\mathrm{ms}$ in our model, while
their estimate is $\sim1000\,\mathrm{ms}$ (see Fig.~4 in their paper).
Note that even our estimate will be a bit over estimation, since the
event horizon is located somewhat farther out from apparent horizon.








Now we move on to the comparison of different models to see the initial
mass dependence of the dynamics and neutrino emissions. The qualitative
features mentioned above, such as the split of the inner core or the
increase of $Y_e$, are common to all the models. The central density,
$\rho_c$, at the moment when the apparent horizon formed is smaller for
more massive models, which is similar the results by Linke et al.
(2001). For instance, we find
$\rho_c=7.51\times10^{14}\,\mathrm{g/cm^3}$ for the reference model
($M_i=375M_\odot$) while $\rho_c=1.23\times10^{14}\,\mathrm{g/cm^3}$ for
the model with $M_i=10500M_\odot$.

\begin{figure}
\plotone{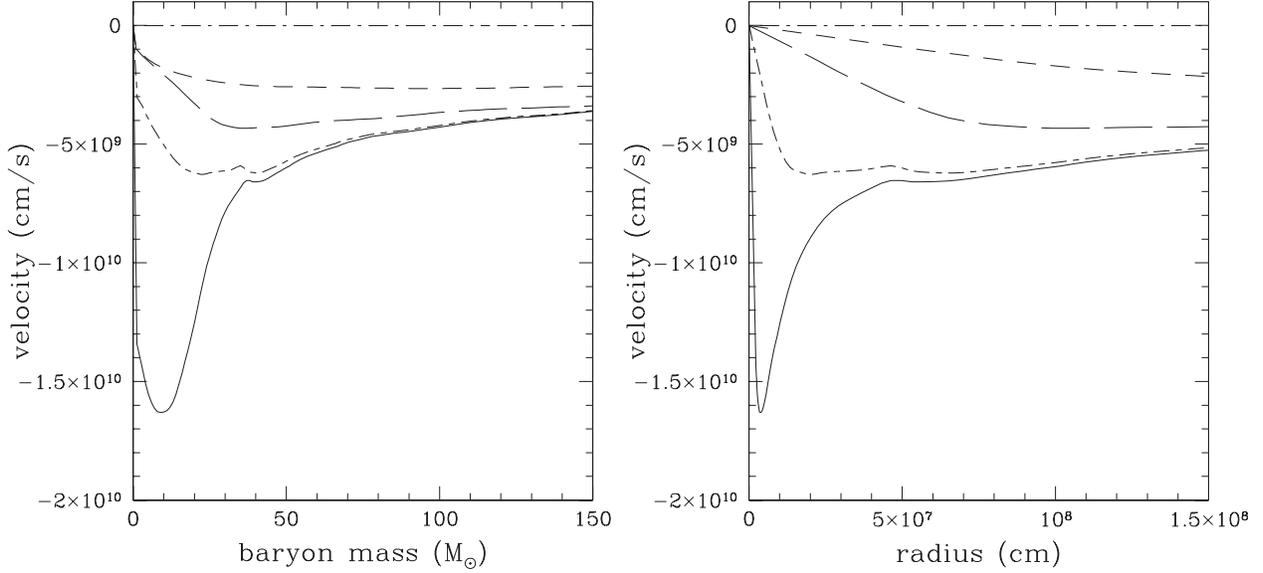}
\caption{Profiles of the radial velocity for the model with
 $M_i=10500M_\odot$. The line correspond, from top to bottom, to
 $t=-11.03\,\mathrm{s}$, $t=-59.8\,\mathrm{ms}$, $t=-14.0\,\mathrm{ms}$,
 $t=-1.27\,\mathrm{ms}$ and $t=0\,\mathrm{ms}$.}
\label{velprf17}
\end{figure}

As the initial mass $M_i$ gets larger, the initial value of entropy per
baryon in the iron core, $s_\mathrm{iron}$, becomes grater. Our results
show that the final value of entropy per baryon at the center of core,
$s_\mathrm{core}$, does not become larger proportionately. For
massive models, in fact, $s_\mathrm{core}$ is saturated. The reason is
that the neutrino cooling is more efficient for more massive stars,
since electrons are non-degenerate and a large amount of
electron-positron pairs exist. For instance, the ratio of the net
electron number to the sum of the numbers of electron and positron at
the center initially is $27.6\%$ for the reference model while it is
$5.2\%$ for the model with $M_i=10500M_\odot$. The existence of a large
amount of electron-positron pairs leads to a smaller difference of the
electron- and positron-captures. This is reflected on the minimum value
of $Y_e$ at the center, which is $0.197$ in the for reference model and
$0.235$ in the model with $M_i=10500M_\odot$.
Owing to the very efficient neutrino cooling, the
inner core mass $M_\mathrm{core}$ and the location of the apparent
horizon $M_\mathrm{AH}$ at $t=0\,\mathrm{ms}$ do not increase with the
initial mass $M_i$, either. These results are summarized in
Table~\ref{m-s_result}. For very massive stars, the inner core fraction
gets smaller and the general relativistic strong gravity pulls not only
the vicinity of the center but also the entire inner core. This makes the
substructure of the inner core indiscernible by $t=0\,\mathrm{ms}$ (for
example, the model with $M_i=10500M_\odot$ ; Fig.~\ref{velprf17}).

The radius of the inner core, which is $\sim3\times10^7\,\mathrm{cm}$
for the reference model and $\sim5\times10^7\,\mathrm{cm}$ for the
model with $M_i=10500M_\odot$, is also insensitive to the initial mass,
$M_i$. However, the radius of the neutrino sphere depends on $M_i$. For
instance, The radius of the neutrino sphere for $\nu_e$ with
$15.8\,\mathrm{MeV}$ at $t=0\,\mathrm{ms}$ is
$\sim1.5\times10^8\,\mathrm{cm}$ for the reference model with
$M_i=375M_\odot$ while it is $\sim10^9\,\mathrm{cm}$ for the model with
$M_i=10500M_\odot$. This is mainly because more massive stars have
thicker outer cores, which are opaque to neutrinos.

As mentioned already, we compute the collapse until after the apparent
horizon is formed inside the inner core. Although it is likely, we do
not know if the event horizon is also inside the inner core, since the
null geodesics should be integrated to infinity, which is impossible. If
this is the case, the black hole formation is not reflected in the
neutrino signal until the inner core is swallowed into the event horizon
because the inner core is the main source of neutrinos as mentioned
earlier. Unfortunately, the numerical difficulty does not allow us to
compute the dynamics up to the point at which the apparent horizon
swallows the inner core entirely, the sufficient condition that the
event horizon is outside the inner core. From the extrapolation of the
previous evolutions we expect that it will take the apparent horizon
$\sim1\,\mathrm{ms}$ to reach the inner core for all models. The
neutrino emission from the inner core is estimated to be negligible in
this phase. Hence, we have to somehow estimate the neutrino emissions
diffusing out of the region between the inner core surface and the
neutrino sphere. We can obtain the upper and lower limits as
follows. The upper limit is estimated assuming that all neutrinos in
this region flow out without absorbed or scattered. For the lower limit,
on the other hand, we assume that all neutrinos in this region are
trapped and do not come out. We summarize the average and total energies
of emitted neutrinos estimated in this way for all the models in
Table~\ref{nutr-eg}.

\begin{figure}
\plotone{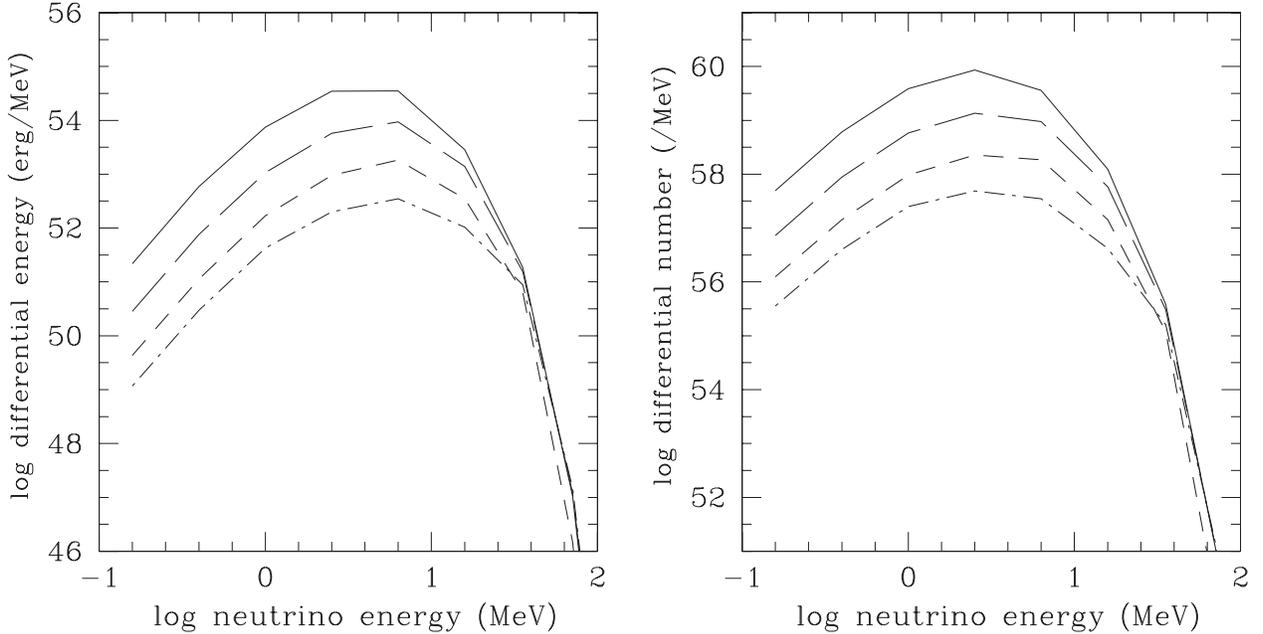}
\caption{Neutrino energy (left) and number (right) spectra. The
 dot-dashed line, short dashed line, long dashed line and solid line
 represent, respectively, the models with  $M_i=375M_\odot$, 
 $M_i=1145M_\odot$, $M_i=3500M_\odot$ and $M_i=10500M_\odot$.}
\label{spems}
\end{figure}

In Fig.~\ref{spems}, we plot the time-integrated energy and number
spectra of neutrinos for 4 different models based on the upper limit
discussed above. As expected, the more massive the initial mass is, the
larger the total emission energy is, since the liberated gravitational
energy is greater. It is interesting, however, that the spectra do not
become harder but rather softer as the initial mass increases. There are
two reasons. The first reason is again that the neutrino cooling is more
efficient for more massive stars, and as a result that, the physical
condition of the inner core, which is a main source of neutrinos, is
similar among different models. The second reason, which is also
mentioned already, is that massive models have a very thick outer core
which prevents high energy neutrinos from getting out of the inner core
directly. As seen in Table~\ref{nutr-eg}, this is particularly
remarkable for $\nu_x$, the species with the lowest reaction rates. The
reason is as follows. The distance from the stellar surface down to the
neutrino sphere is almost independent of the initial mass. On the other
hand, the distance from the surface of the inner core, the main neutrino
source, to the stellar surface is dependent on it. As the initial mass
increases, so does the distance between the inner core surface and the
neutrino sphere. Since the neutrino sphere for $\nu_x$ is inside of those
for $\nu_e$ and $\bar\nu_e$ and the temperature scale height is smaller
near the inner core, the average energy of $\nu_x$ is more sensitive to
the difference of the initial mass. 



\section{Relic Neutrinos from Pop III Massive Stars} \label{neutrino}

In this section, we estimate the number flux of diffuse relic
neutrinos from Pop III massive stars. After giving the formulation,
we apply it to several star formation histories, and discuss the
possibility to constrain them. In the following, we do not take into
account the mixing of neutrinos for simplicity.

\subsection{Formulation for Relic Neutrino Background} 

Here we formulate the number flux of the relic neutrino from Pop III
massive stars. We assume that Pop III stars in the mass interval
$M_0\leq m \leq M_N$ collapse and emit neutrinos at redshift 
$z_i \geq z\geq z_f$. Then the present number flux of relic
neutrinos on the earth is given by 
\begin{equation}
\frac{dF_\nu}{dE_\nu} = c \int^{z_f}_{z_i} \int^{M_N}_{M_0} \frac{dN(m,E^\prime_\nu)}{dE^\prime_\nu}\,(1+z)\,R_\mathrm{Pop III}(z,m)\,dm\,\frac{dt}{dz}\,dz,
\label{flux_def}
\end{equation}
where $E_\nu$ is the detected neutrino energy and
$E^\prime_\nu=(1+z)E_\nu$ is the emitted neutrino energy. The neutrino
number spectrum emitted by the progenitor with mass $m$ is
$dN(m,E^\prime_\nu)/dE^\prime_\nu$. We denote the birth rate of Pop III
massive stars per comoving volume per mass as $R_\mathrm{PopIII}(z,m)$.
Massive stars are supposed to die immediately after the birth. The
relation between $t$ and $z$ is given by  
\begin{equation}
\frac{dz}{dt} = -H_0 (1+z) \sqrt{\Omega_m(1+z)^3 + \Omega_\Lambda},
\label{dzdt}
\end{equation}
and according to the standard $\Lambda$CDM cosmology, the cosmological
parameters are given as $\Omega_m = 0.3$, $\Omega_\Lambda = 0.7$ and 
$H_0=71^{+4}_{-3}\,\mathrm{km/s/Mpc}$ (Spergel et al. 2003). 

Here we assume that $R_\mathrm{Pop III}(z,m)$ can be written as
\begin{equation}
R_\mathrm{Pop III}(z,m)\,dm\,dz = \frac{dn(m)}{dm}\, dm\,\psi(z)\,dz = dn(m)\,\psi(z)\,dz,
\label{divide}
\end{equation}
where $dn(m)$ is the number of Pop III massive stars within the mass
interval $[m,m+dm]$. The normalization factor $\psi(z)$ is chosen as
\begin{equation}
\int^{z_f}_{z_i} \psi(z) \frac{dt}{dz} dz = \int^{z_i}_{z_f} \frac{\psi(z)}{H_0(1+z)\sqrt{\Omega_m(1+z)^3 + \Omega_\Lambda}}dz = 1,
\label{psinorm}
\end{equation}
Thus we can rewrite eq. (\ref{flux_def}) as
\begin{equation}
\frac{dF_\nu}{dE_\nu} = c \int^{M_N}_{M_0} dm \frac{dn(m)}{dm} \int^{z_i}_{z_f} dz \frac{\psi(z)}{H_0(1+z)\sqrt{\Omega_m(1+z)^3 + \Omega_\Lambda}} \frac{dN(m,E^\prime_\nu)}{dE^\prime_\nu} (1+z) .
\label{flux_mod}
\end{equation}

In the above equation, $dn(m)/dm$ represents the initial mass function
(IMF). Here we adopt the IMF of Pop III stars proposed by Nakamura \&
Umemura (2001). Their IMF is bimodal and the heavier component is given as 
\begin{equation}
\left\{
\begin{array}{lr}
\displaystyle{\frac{dn}{dm}} = Bm^{-\beta-1} & \mathrm{for}\quad m \geq M_\mathrm{min}, \\ 
n = 0 & \mathrm{for}\quad m < M_\mathrm{min}, \\ 
\end{array}
\right.
\label{nu_imf}
\end{equation}
where $\beta>1$ and $B>0$ are independent of $m$. Incidentally, the IMF
given by Salpeter (1955) has $\beta=1.35$ for the mass range,
$0.4M_\odot<m<10M_\odot$. The mass density of Pop III massive stars is
given as
\begin{equation}
P_\mathrm{all} = \int^{\infty}_{M_\mathrm{min}} m dn = \frac{B}{\beta-1} {M_\mathrm{min}}^{1-\beta}.
\label{mall_int}
\end{equation}
On the other hand, we denote the time-integrated fraction of cosmic
baryon converted into Pop III stars as $\epsilon$ and the fraction
of the mass contained in the heavier population as $(1-\kappa)$. Then
$P_\mathrm{all}$ can be also expressed as 
\begin{equation}
P_\mathrm{all} = n_b\, m_N\, \epsilon\, (1-\kappa),
\label{mall_prd}
\end{equation}
where $n_b=(2.5\pm0.1)\times10^{-7}\,\mathrm{cm^{-3}}$ is the
present number density of cosmic baryon (Spergel et al. 2003) and $m_N$
is the nucleon mass. From eqs. (\ref{mall_int}) and (\ref{mall_prd}),
we can determine the normalization factor $B$ as 
\begin{equation}
B = (\beta-1)\,{M_\mathrm{min}}^{\beta-1}\,n_b\, m_N\, \epsilon\, (1-\kappa).
\label{mass_bin}
\end{equation}

In order to evaluate the relic neutrino flux from our models, we
approximate the integral in eq. (\ref{flux_mod}) by the summation over
mass bins covering the mass range of our models. Using the equation 
\begin{equation}
\int^{M_k}_{M_{k-1}} dm \frac{dn(m)}{dm} = \frac{\beta-1}{\beta}\, n_b\, m_N\, \epsilon\, (1-\kappa)\,{M_\mathrm{min}}^{\beta-1}\displaystyle\left({M_{k-1}}^{-\beta}-{M_k}^{-\beta}\right),
\label{prm_B}
\end{equation}
we rewrite eq. (\ref{flux_mod}) as
\begin{equation}
\begin{array}{ll}
\displaystyle \frac{dF_\nu}{dE_\nu} = & \displaystyle  \frac{\beta-1}{\beta}\, c\, n_b\, m_N\, \epsilon\, (1-\kappa)\,{M_\mathrm{min}}^{\beta-1} \sum^N_{k=1} \left({M_{k-1}}^{-\beta}-{M_k}^{-\beta}\right) \\
 & \times \displaystyle \int^{z_i}_{z_f} dz \frac{\psi(z)}{H_0(1+z)\sqrt{\Omega_m(1+z)^3 + \Omega_\Lambda}} \frac{dN(\widetilde{M_k},E^\prime_\nu)}{dE^\prime_\nu} (1+z) , \\
\end{array}
\label{flux_dis}
\end{equation}
where $\widetilde{M_k}$ is approximated by the geometric mean of
$M_{k-1}$ and $M_k$, and is chosen to coincide with the mass of our
models. 
$M_k$ is independent of $\beta$ and given in Table~\ref{mass_bin_c}.
This  approximation is estimated to give at most several percents of
error for the range of $1<\beta<3$. Here we use 18 mass bins starting
from $M_0=260M_\odot$, the largest mass for the pair-instability
supernova, to $M_{18}=16000M_\odot$, the smallest mass for the onset of
collapse by the general relativistic effect. 

In the following, we take the most optimistic values for the uncertain
quantities in estimating the relic neutrinos. For the total neutrino
emissions, we take the upper limit given in Table~\ref{nutr-eg}. In the
IMF by Nakamura \& Umemura (2001), we choose $(1-\kappa)=1$. Since they
estimate that $M_\mathrm{min}$ is in the range of a few times
$10\,-\,100M_\odot$, we take $M_\mathrm{min}=100M_\odot$. The Pop III
star formation efficiency, $\epsilon$, one of the most uncertain
factors, is suggested to be rather large, of the order of $10\%$, from
the estimations of the contribution of Pop III stars
to the cosmic infrared background (e.g., Santos, Bromm, \& Kamionkowski
2002 ; Salvaterra, \& Ferrara 2003 ; However, see also Madau, \& Silk
2005). Here we assume $\epsilon = 0.1$. The change of the relic neutrino
flux due to the variations of $(1-\kappa)$, $M_\mathrm{min}$ or $\epsilon$
is obtained from eq. (\ref{flux_dis}). We find that these parameters do
not affect the peak energy of neutrino. $\beta$ is also an uncertain
parameter, and its influence on the flux is complicated. Hence we take
several values.

As for the star formation history, $\psi(z)$, we employ the following
three models. The observation by WMAP suggests that the reionization
occurred at redshift $z=17\pm5$ (Spergel et al. 2003). Based on this, we
assume the following function for $\psi(z)$ as model A.
\begin{equation}
\psi(z) = \psi_\mathrm{A}(z) \equiv \frac{1}{5\sqrt{2\pi}} \exp \displaystyle\left(-\frac{(z-17)^2}{20}\right) \,H_0\,(1+z)\,\sqrt{\Omega_m(1+z)^3 + \Omega_\Lambda}.
\label{case-a}
\end{equation}
On the other hand, according to the theoretical investigation of the
star formation history by Scannapieco, Schneider, \& Ferrara (2003), the
peak of the Pop III star formation might have been at $z \sim 10$.
Hence, as model B,
\begin{equation}
\psi(z) = \psi_\mathrm{B}(z) \equiv \delta(z-10)\,H_0\,(1+z)\,\sqrt{\Omega_m(1+z)^3 + \Omega_\Lambda}.
\label{case-b}
\end{equation}
Some authors attempt to estimate the Pop III star formation rate using
the gamma ray bursts (Yonetoku et al. 2004 ; Murakami et al. 2005), and
suggest a rather continuous formation of Pop III stars. Based on their
results, $\psi(z)$ is assumed to be 
\begin{equation}
\psi(z) = \psi_\mathrm{C}(z) \propto (1+z)^{1.7} \qquad \mathrm{for} \quad 4 < z < 12,
\label{case-c}
\end{equation}
in model C.

\subsection{Results and Discussion}

\begin{figure}
\plotone{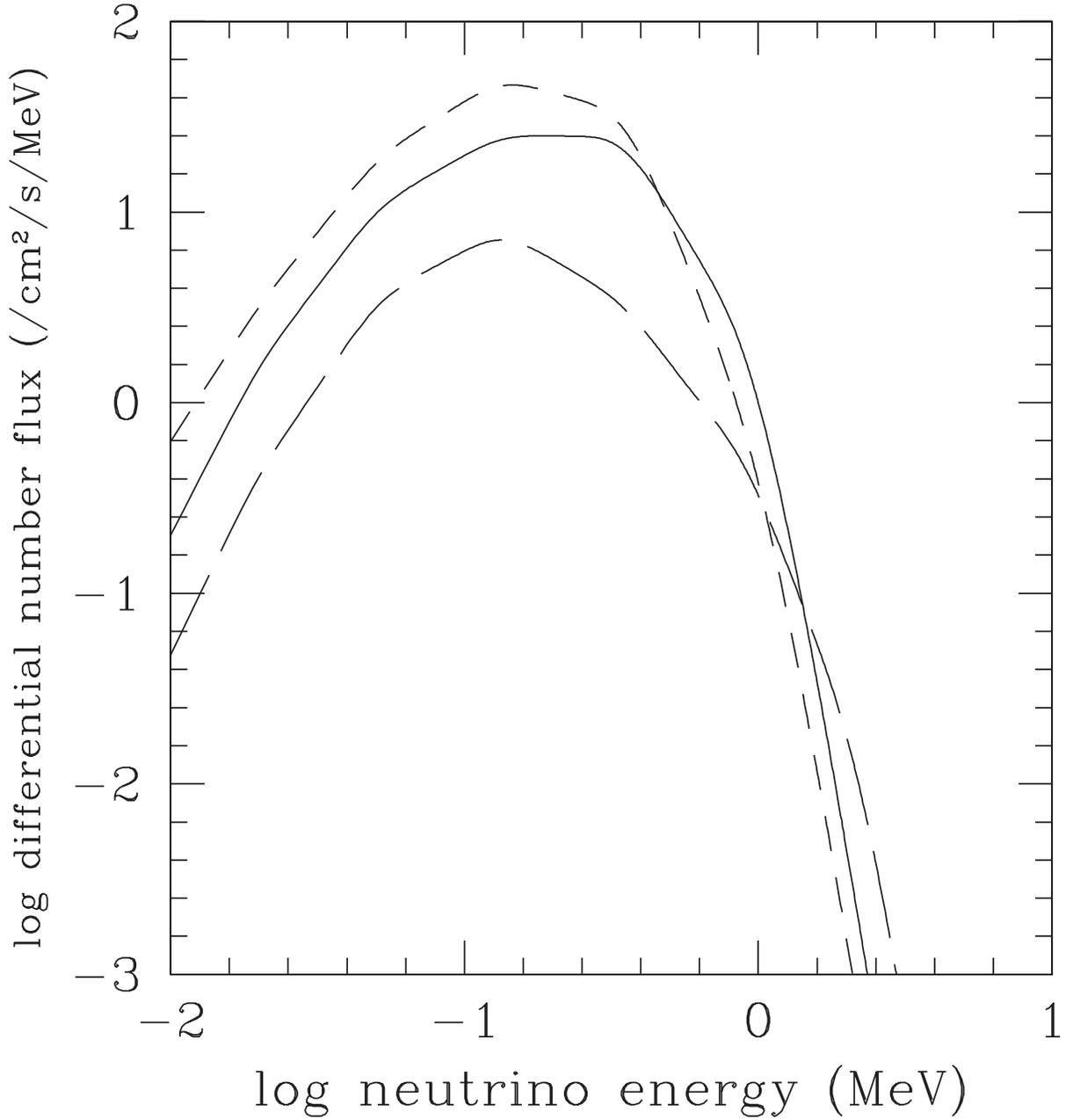}
\caption{Relic neutrino number flux from Pop III massive stars (model A,
 $\beta = 1.35$). The short dashed line, solid line and long
 dashed line represent $\nu_e$, $\bar \nu_e$ and $\nu_x$, respectively.}
\label{relic_stdm}
\end{figure}

In Fig.~\ref{relic_stdm}, we plot the relic neutrino fluxes for model
A and $\beta=1.35$. It is seen that the peak energy of $\nu_x$ ($=$
$\nu_\mu$, $\nu_\tau$, $\bar \nu_\mu$, $\bar \nu_\tau$) is smaller than
$\nu_e$ and $\bar \nu_e$. At first glance, this seems inconsistent with
the trend of the average energy shown in Table~\ref{nutr-eg}. However,
the number flux of $\nu_x$ has a high energy tail because of the absence
of charged current reactions. Integrated over the spectrum, the mean
energy of $\nu_x$ becomes higher than those of $\nu_e$ and $\bar \nu_e$
as in Table~\ref{nutr-eg}.

\begin{figure}
\plotone{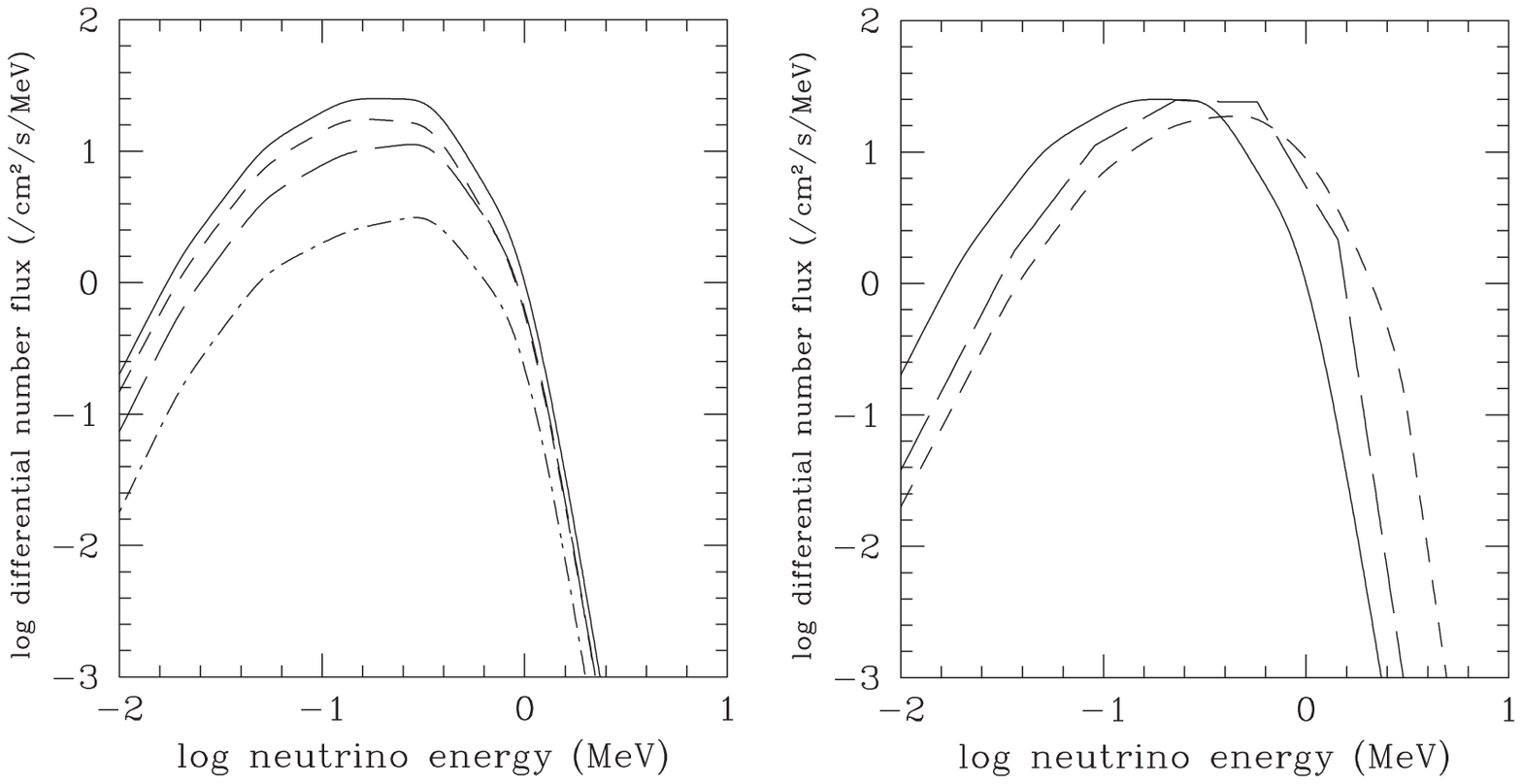}
\caption{Relic $\bar \nu_e$ number fluxes from Pop III massive stars for
 various values of $\beta$ and different star formation histories. The left
 panel shows the result of model A and the short dashed line, solid
 line, long dashed line and dot-dashed 
 line correspond to $\beta = 1.1$, $\beta = 1.35$, $\beta = 2$ and
 $\beta = 3$, respectively. The right panel shows the result for $\beta = 1.35$ and the solid line, long dashed line, and short dashed line represent models A, B and C, respectively.}
\label{relic_vrsm}
\end{figure}

We show in the left panel of Fig.~\ref{relic_vrsm} the relic $\bar\nu_e$
fluxes for different values of $\beta$. We find that the peak energy is
not sensitive to $\beta$. This is due to the fact that the mean energy
of the neutrinos is insensitive to the initial stellar mass, which we
have already mentioned. We also find that the model with $\beta=1.35$
gives the greatest number flux. This can be understood as follows. We
first remind readers that the total mass density, $P_\mathrm{all}$, is
fixed when we vary $\beta$ (see eqs. (\ref{mall_int}) and
(\ref{mall_prd})). If $\beta$ is close to unity, a substantial fraction
of stars falls in the realm of supermassive stars
($\gtrsim16000M_\odot$), which we ignore in this paper. The same is true
for larger $\beta$. In this case, a large population of stars contribute
to the pair-instability supernovae ($\lesssim 260M_\odot$), which are
not considered here, either. In the right panel of
Fig.~\ref{relic_vrsm}, we compare 
different star formation histories. It is found that the larger the
redshift at which Pop III massive stars are formed is, the lower the
peak energy becomes. We have seen that the peak energy is not sensitive
to IMF. Hence the peak energy is determined solely by the redshift of
the Pop III star formation.

As for the possibilities of detection, we must say that it is difficult
for the currently operating detectors. In spite of the difference of
our results and what is found by Iocco et al. (2005) based on the
results by Fryer, Woosley, \& Heger (2001), both results are negative
for detection, because the cosmological redshift reduces the peak energy
to lower values. To put it more precisely, the relic $\nu_e$ fluxes from
Pop III massive 
stars are overwhelmed by solar $\nu_e$ below $18\,\mathrm{MeV}$ and
relic neutrinos by ordinary supernova above $\sim10\,\mathrm{MeV}$. As
for $\bar \nu_e$, the emissions from nuclear reactors are the main
obstacle below $10\,\mathrm{MeV}$. Thus the existing detectors can
not distinguish Pop III relic neutrinos from others. However, because
the solar and reactor neutrinos are not isotropic, removing them is
possible at least in principle. For $\bar \nu_e$, in particular, Pop III
massive stars are the largest cosmological sources. In the future, we
may be able to discuss the Pop III star formation history with these
diffuse neutrino fluxes.




\section{Conclusions}

In this paper, we have numerically studied the spherical gravitational
collapse of the Pop III massive stars, taking into account the reactions
and transports of neutrinos in detail. Neutrinos affect the dynamics of
collapse crucially, determining the inner core mass, the location and
formation time of the apparent horizon. Moreover, it gives rise to the
non-monotonic evolution of the electron fraction during the collapse,
which is not seen in the ordinary supernova core. Neutrino cooling is
very efficient and the final value of core entropy is not sensitive to
the initial value or the initial stellar mass. For very massive stars,
even the outer core becomes opaque to neutrinos. As a result, the
neutrino spectra do not become harder as the initial mass
increases. Therefore, the peak energy of relic neutrinos is mainly
determined by the redshift of the Pop III star formation and not
sensitive to the IMF. At present, the detection of these relic neutrinos
is highly difficult.

These Pop III massive stars might have been rotating rapidly and have
had magnetic fields, both of which we ignored in this paper. Although
the numerical treatment of general relativity and neutrino transport in
multi-dimension is a challenging problem, rapidly rotating Pop III
massive stars with magnetic fields are proposed to be associated
with gamma ray bursts (e.g., Schneider et al. 2002), and it is certainly
worth investigation.

\acknowledgments

One of the authors (K.N.) is grateful to Hideki Maeda for valuable
discussions. He would like to thank Hideyuki Suzuki for useful
comments and aids in computations.
In this work, numerical computations were partially performed on Fujitsu
VPP5000 at the Astronomical Data Analysis Center, ADAC, of the National
Astronomical Observatory of Japan (VPP5000 System Projects rkn60c,
wkn10b), and on the supercomputers in RIKEN and KEK (KEK Supercomputer
Project No.~108).
This work was partially supported by Grants-in-Aid for the Scientific
Research from the Ministry of Education, Science and Culture of Japan
through No.14079202, No.14740166, No.15740160, and The 21st century COE Program
``Holistic Research and Education Center for Physics of Self-organization
Systems''. 






\appendix

\clearpage

\begin{table}
\caption{Correspondence of the Initial Mass to the Iron Core Entropy}
\begin{center}
\begin{tabular}{rrrrrrrrr}
\tableline\tableline
\multicolumn{1}{c}{$M_i$} & $M_\mathrm{He}$ & $M_\mathrm{O}$ & $s_\mathrm{O}$ & $s_\mathrm{iron}$ \\ \hline
  300 &  159 &  143 & 14.54 & 15.98 \\
  375 &  201 &  181 & 16.06 & 17.50 \\
  470 &  254 &  228 & 17.73 & 19.17 \\ 
  585 &  319 &  287 & 19.53 & 20.96 \\
  730 &  400 &  360 & 21.54 & 22.97 \\
  915 &  504 &  454 & 23.81 & 25.25 \\ 
 1145 &  633 &  570 & 26.33 & 27.77 \\
 1430 &  794 &  714 & 29.10 & 30.54 \\
 1800 & 1001 &  901 & 32.29 & 33.74 \\
 2250 & 1254 & 1129 & 35.74 & 37.19 \\
 2800 & 1563 & 1407 & 39.51 & 40.96 \\
 3500 & 1956 & 1760 & 43.79 & 45.24 \\ 
 4350 & 2434 & 2191 & 48.44 & 49.89 \\
 5500 & 3080 & 2772 & 54.04 & 55.50 \\
 6800 & 3810 & 3429 & 59.70 & 61.16 \\ 
 8500 & 4765 & 4288 & 66.32 & 67.78 \\
10500 & 5889 & 5300 & 73.29 & 74.75 \\
13500 & 7574 & 6817 & 82.59 & 84.06 \\ \hline
\end{tabular}
\end{center}
\label{m-s_relation}
\tablecomments{$M_i$, $M_\mathrm{He}$ and $M_\mathrm{O}$ denote the
 initial stellar mass, helium core mass and oxygen core mass
 respectively, and they are in the unit of the solar mass
 ($M_\odot$). The entropy par baryon in the oxygen core and the iron
 core is denoted as $s_\mathrm{O}$ and $s_\mathrm{iron}$, respectively,
 and they are in the unit of the Boltzmann constant ($k_\mathrm{B}$).}
\end{table}

\begin{table}
\caption{Initial Mass Dependence of Core Entropy, Core Mass and Location of Apparent Horizon.}
\begin{center}
\begin{tabular}{rrrrrrrrr}
\tableline\tableline
\multicolumn{1}{c}{$M_i$} &  $s_\mathrm{iron}$ & $s_\mathrm{core}$ &
 $M_\mathrm{core}$ & $M_\mathrm{AH}$ \\ \hline
  300 & 15.98 &  7.33 & 13.6 & 4.17 \\
  375 & 17.50 &  7.37 & 14.1 & 4.08 \\
  470 & 19.17 &  7.70 & 15.8 & 4.53 \\ 
  585 & 20.96 &  7.88 & 16.8 & 4.84 \\
  730 & 22.97 &  8.07 & 18.2 & 5.18 \\
  915 & 25.25 &  8.28 & 19.5 & 5.39 \\ 
 1145 & 27.77 &  8.50 & 21.0 & 5.82 \\
 1430 & 30.54 &  8.77 & 23.4 & 6.33 \\
 1800 & 33.74 &  8.90 & 24.7 & 6.46 \\
 2250 & 37.19 &  9.24 & 26.5 & 6.96 \\
 2800 & 40.96 &  9.64 & 30.0 & 7.60 \\
 3500 & 45.24 & 10.10 & 33.5 & 8.42 \\ 
 4350 & 49.89 & 10.35 & 36.0 & 8.44 \\
 5500 & 55.50 & 10.56 & 39.0 & 9.11 \\
 6800 & 61.16 & 10.41 & 37.6 & 9.10 \\ 
 8500 & 67.78 & 10.58 & 40.1 & 9.40 \\
10500 & 74.75 & 10.48 & 39.5 & 9.26 \\
13500 & 84.06 & 11.20 & 44.5 & 9.91 \\ \hline
\end{tabular}
\end{center}
\label{m-s_result}
\tablecomments{$M_i$ and $s_\mathrm{iron}$ are the same as in
 Table~\ref{m-s_relation}. $s_\mathrm{iron}$ is the entropy
 par baryon at the center of the inner core at $t=0$, and it is
 in the unit of the Boltzmann constant ($k_\mathrm{B}$).
 $M_\mathrm{core}$ is the mass of the inner core at $t=0$ and
 $M_\mathrm{AH}$ is the location of the apparent horizon at $t=0$. They
 are in the unit of the solar mass ($M_\odot$).}
\end{table}

\begin{deluxetable}{cccccccc}
\tabletypesize{\scriptsize}
\tablewidth{0pt}
\tablecaption{Estimates of Average and Total Energies of Emitted Neutrinos.}
\tablehead{$M_i$ & $\langle{E_{\nu_e}\rangle}$ &
 $\langle{E_{\bar \nu_e}\rangle}$ & $\langle{E_{\nu_x}\rangle}$ & 
 $E^\mathrm{tot}_{\nu_e,52}$ & $E^\mathrm{tot}_{\bar \nu_e,52}$ &
 $E^\mathrm{tot}_{\nu_x,52}$ & $E^\mathrm{tot}_{\mathrm{all},52}$}
\startdata
  300 & 5.30 - 5.33 & 8.00 - 8.04 & 10.21 - 10.89 & 12.77 - 12.87 &  9.58 -  9.69 &  2.54 -  2.78 & 32.49 - 33.69 \\
  375 & 5.06 - 5.10 & 7.20 - 7.28 &  8.26 -  8.98 & 17.16 - 17.38 & 12.81 - 13.10 &  3.04 -  3.41 & 42.14 - 44.10 \\
  470 & 5.27 - 5.33 & 7.57 - 7.68 &  8.46 -  9.09 & 23.83 - 24.21 & 18.80 - 19.35 &  3.58 -  3.93 & 56.90 - 59.27 \\ 
  585 & 5.22 - 5.28 & 7.36 - 7.45 &  7.66 -  8.11 & 31.31 - 31.82 & 25.94 - 26.60 &  4.38 -  4.72 & 74.76 - 77.30 \\
  730 & 5.18 - 5.24 & 7.16 - 7.25 &  7.10 -  7.43 & 41.90 - 42.62 & 35.75 - 36.67 &  5.69 -  5.96 & 100.1 - 103.1 \\
  915 & 5.13 - 5.20 & 6.98 - 7.07 &  6.66 -  6.90 & 56.39 - 57.48 & 49.41 - 50.79 &  7.50 -  7.84 & 135.8 - 139.6 \\ 
 1145 & 5.08 - 5.16 & 6.79 - 6.89 &  6.35 -  6.53 & 75.57 - 77.21 & 67.71 - 69.67 & 10.29 - 10.66 & 184.4 - 189.5 \\
 1430 & 5.03 - 5.11 & 6.62 - 6.72 &  6.05 -  6.19 & 101.7 - 103.9 & 92.64 - 95.36 & 14.20 - 14.59 & 251.1 - 257.6 \\
 1800 & 5.01 - 5.09 & 6.52 - 6.62 &  6.03 -  6.20 & 135.0 - 138.5 & 127.6 - 131.5 & 20.02 - 20.82 & 342.7 - 353.3 \\
 2250 & 4.88 - 4.97 & 6.21 - 6.32 &  5.66 -  5.83 & 182.3 - 187.4 & 171.2 - 177.2 & 28.63 - 29.88 & 468.1 - 484.1 \\
 2800 & 4.89 - 4.99 & 6.16 - 6.28 &  5.69 -  5.90 & 248.3 - 256.2 & 236.6 - 245.3 & 41.60 - 43.96 & 651.3 - 677.3 \\
 3500 & 4.83 - 4.93 & 6.00 - 6.14 &  5.60 -  5.81 & 330.3 - 341.8 & 320.6 - 333.9 & 59.91 - 63.44 & 890.6 - 929.5 \\ 
 4350 & 4.73 - 4.83 & 5.83 - 5.95 &  5.44 -  5.63 & 422.5 - 437.5 & 424.7 - 441.3 & 82.79 - 87.36 &  1178 -  1228 \\
 5500 & 4.63 - 4.74 & 5.64 - 5.76 &  5.29 -  5.46 & 562.3 - 582.2 & 575.5 - 596.7 & 117.2 - 123.4 &  1608 -  1672 \\
 6800 & 4.47 - 4.55 & 5.36 - 5.46 &  5.00 -  5.11 & 695.2 - 715.1 & 728.3 - 749.7 & 148.6 - 163.8 &  2018 -  2194 \\ 
 8500 & 4.35 - 4.43 & 5.15 - 5.23 &  4.58 -  4.67 & 899.2 - 924.6 & 952.3 - 979.1 & 202.8 - 209.4 &  2663 -  2741 \\
10500 & 4.05 - 4.12 & 4.92 - 4.99 &  4.31 -  4.38 &  1124 -  1152 &  1204 -  1233 & 261.5 - 267.5 &  3374 -  3455 \\
13500 & 3.94 - 4.00 & 4.75 - 4.82 &  4.20 -  4.28 &  1558 -  1597 &
 1669 -  1712 & 388.2 - 398.4 &  4780 -  4902 \\
\enddata
\label{nutr-eg}
\tablecomments{The mean energy of emitted $\nu_i$ is denoted as
 $\langle{E_{\nu_i}\rangle} \equiv E^\mathrm{tot}_{\nu_i}/N^\mathrm{tot}_{\nu_i}$ and the lower and upper limits of them are given in the unit of MeV, where $E^\mathrm{tot}_{\nu_i}$ and $N^\mathrm{tot}_{\nu_i}$ are the total energy and number of neutrinos. The subscript ``52'' means the quantity given in the unit of $10^{52}\mathrm{erg}$. $E^\mathrm{tot}_{\mathrm{all}}$ is the total energy summed over all species.}
\end{deluxetable}

\begin{table}
\caption{Binning of IMF}
\begin{center}
\begin{tabular}{rrrrrrrrrrrrrrrrrrrr}
\tableline\tableline
\multicolumn{1}{c}{$k$} & 0 & 1 & 2 & 3 & 4 & 5 & 6 & 7 & 8 & 9 \\ \hline
$M_k$ & 260 & 335 & 420 & 525 & 655 & 815 & 1025 & 1280 & 1600 & 2010 \\ \hline \hline
\multicolumn{1}{c}{$k$} & 10 & 11 & 12 & 13 & 14 & 15 & 16 & 17 & 18 \\ \hline
$M_k$ & 2500 & 3150 & 3900 & 4900 & 6100 & 7600 & 9450 & 11650 & 16000 \\ \hline
\end{tabular}
\end{center}
\label{mass_bin_c}
\end{table}


\begin{thebibliography}{}
\bibitem[]{}
 Ando, S., \& Sato, K. 2004, New J. Phys., 6, 170
\bibitem[]{}
 Bond, J.R., Arnett, W.D., \& Carr, B.J. 1984, \apj, 280, 825
\bibitem[]{}
 Bruenn, S.W. 1985, \apjs, 58, 771
\bibitem[]{}
 Frebel, A., et al. 2005, Nature, 434, 871  
\bibitem[]{}
 Fryer, C.L., Woosley, S.E., \& Heger, A. 2001, \apj, 550, 372  
\bibitem[]{}
 Herant, M., Benz, W., Hix, W.R., Fryer, C.L., \& Colgate, C.A. 1994, \apj, 435, 339 
\bibitem[]{}
 Iocco, F., Mangano, G., Miele, G., Raffelt, G.G., \& Serpico, P.D. 2005, Astropart. Phys., 23, 303 
\bibitem[]{}
 Linke, F., Font, J.A., Janka, H.-Th, M$\ddot{\mathrm{u}}$ller, E., \& Papadopoulos, P. 2001, A\&A, 376, 568
\bibitem[]{}
 Madau, P., \& Silk, J. 2005, MNRAS, 359, L37  
\bibitem[]{}
 Maillard, J.P., Paumard, T., Stolovy, S.R., \& Riguaut, F. 2004, A\&A, 423, 155  
\bibitem[]{}
 Misner, C.W., \& Sharp, D.H. 1964, Phys. Rev., 136, 571  
\bibitem[]{}
 Murakami, T., et al. 2005, \apj, 625, L13  
\bibitem[]{}
 Nakamura, F., \& Umemura, M. 2001, \apj, 548, 19  
\bibitem[]{}
 Salpeter, E.E. 1955, \apj, 122, 161
\bibitem[]{}
 Salvaterra, R., \& Ferrara, A. 2003, MNRAS, 339, 973  
\bibitem[]{}
 Santos, M.R., Bromm, V., \& Kamionkowski, M. 2002, MNRAS, 336, 1082  
\bibitem[]{}
 Scannapieco, E., Schneider, R., \& Ferrara, A. 2003, \apj, 589, 35
\bibitem[]{}
 Schneider, R., Ferrara, A., Natarajan, P., \& Omukai, K. 2002, \apj, 571, 30
\bibitem[]{}
 Shapiro, S.L., \& Teukolsky, S.A. 1983, in Black Holes, White Dwarfs,
 and Neutron Stars (New York:Wiley)
\bibitem[]{}
 Shen, H., Toki, H., Oyamatsu, K., \& Sumiyoshi, K. 1997, Prog. Theor. Phys., 100, 1013
\bibitem[]{}
 Shen, H., Toki, H., Oyamatsu, K., \& Sumiyoshi, K. 1997, Nucl. Phys., A637, 435
\bibitem[]{}
 Suzuki, H. 1990, Ph.D. thesis, Univ. Tokyo
\bibitem[]{}
 Spergel, D.N., et al. 2003, \apjs, 148, 174
\bibitem[]{}
 Strigari, L.E., Kaplinghat, M., Steigman, G., \& Walker, T.P. 2004, J. Cosmol. Astropart. Phys., 0403, 007
\bibitem[]{}
 Sumiyoshi, K., Yamada, S., Suzuki, H., Shen, H., Chiba, S., \& Toki, H. 2005, \apj, in Press.
\bibitem[]{}
 Umeda, H., \& Nomoto, K. 2002, \apj, 565, 385  
\bibitem[]{}
 van Riper K.A. 1979, \apj, 232, 558  
\bibitem[]{}
 Wald, R.M. 1984, in General Relativity (Chicago and London:The University of Chicago Press) 
\bibitem[]{}
 Woosley, S.E., Heger, A., \& Weaver, T.A. 2002, Rev. Mod. Phys., 74, 1015 
\bibitem[]{}
 Yamada, S. 1997, \apj, 475, 720
\bibitem[]{}
 Yamada, S., Janka, H.-Th, \& Suzuki, H. 1999, A\&A, 344, 533
\bibitem[]{}
 Yonetoku, D., et al. 2004, \apj, 609, 935
\end{thebibliography}
\end{document}